\documentclass[11pt, a4paper]{article}

\usepackage{jheppub}
\usepackage{graphicx}
\usepackage{slashed}
\usepackage{color}	
\usepackage{multirow}
\usepackage{amsmath}
\usepackage{amssymb}
\usepackage{bbold}

\title{Narrow Resonances Revisited - Simplifying Multidimensional Constraints}

\author[a]{R. Sekhar Chivukula,}
\author[b]{Pawin Ittisamai,}
\author[a]{James Osborne,}
\author[a]{and Elizabeth H. Simmons}

\affiliation[a]{Department of Physics and Astronomy, 9500 Gilman Drive, University of California, San Diego}
\affiliation[b]{Department of Physics, Faculty of Science, Chulalongkorn University, Bangkok, Thailand}

\emailAdd{rschivukula@physics.ucsd.edu}
\emailAdd{pawin.i@chula.ac.th}
\emailAdd{josborne@physics.ucsd.edu}
\emailAdd{ehsimmons@ucsd.edu}

\abstract{
  As we amass more LHC data, we continue to search for new and improved methods of visualizing search results, in ways that are as model-independent as possible. The simplified limits framework is an approach developed to recast limits on searches for narrow resonances in terms of products of branching ratios (BRs) corresponding to the resonance's production and decay modes. In this work, we extend the simplified limits framework to a multidimensional parameter space of BRs, which can be used to unfold an ambiguity in the simplified parameter $\zeta$ introduced when more than one channel contributes to the production of the resonance. It is also naturally applicable to combining constraints from experimental searches with different observed final states. Constraints can be visualized in a three-dimensional space of branching ratios by employing ternary diagrams, triangle plots which utilize the inherent unitarity of the sum of the resonance's BRs. To demonstrate this new methodology, we recast constraints from recent ATLAS searches in diboson final states for spin-0, 1, and 2 narrow resonances into constraints on the resonance's width-to-mass ratio and display them in the space of relevant branching ratios. We also demonstrate how to generalize the method to cases where more than three branching ratios are relevant by using N-simplex diagrams, and we suggest a broader application of the general method to digital data sets.
}

\notoc
\compress

\begin{document}
  
  \maketitle

  \section{Introduction}
  \label{sec:intro}
  
  Narrow resonance searches have long been a staple of experimental efforts to identify or constrain new physics beyond the standard model (BSM). Typically, the invariant mass of the system is measured and limits are placed on the production cross section times branching ratio, $\sigma_\textrm{prod} \times \textrm{BR}$, of new resonances and interpreted within the context of specific benchmark (BM) models. In refs.~\cite{Chivukula:2016hvp,Chivukula:2017lyk}, the authors explored recasting these model-independent constraints in terms of a simplified parameter, $\zeta$, which depends only on the product of BRs and the ratio of the resonance width to its mass. This reparameterization of narrow resonance limits in terms of partonic quantities can often simplify the task of interpreting constraints in terms of many models of interest.
  
  Combining narrow resonance searches from multiple channels can extend the exclusion limits of current and future searches for BSM physics with more than one dominant decay mode. This possibility has been explored by both ATLAS~\cite{Aaboud:2018bun} and CMS~\cite{Sirunyan:2019vgt} in diboson and dilepton channels at the LHC. The constraints from different channels were combined in the context of specific BM models where the relationships between BRs are known, a necessity in order to cast constraints in terms of $\sigma_\textrm{prod} \times \textrm{BR}$. This, however, introduces an additional level of model dependence to the results, compared to monochannel searches, which makes it difficult to apply such constraints to models whose BRs vary from the BM choices.
  
  In this work, we extend the concept of simplified limits, demonstrating a methodology by which one may display constraints on the masses or widths of narrow resonances within the parameter space of up to three BRs \textit{without introducing new sources of model dependence}. This method can be applied to single channel searches where the production of a BSM resonance is predicted to receive contributions from multiple channels. It can also be readily applied to combined constraints on BSM scenarios with a single production mode and several decay modes.
  
  In sec.~\ref{sec:simpLimits}, we review the foundations of simplified limits for a single experimental search. In sec.~\ref{sec:ternary}, we introduce ternary diagrams as a method of unfolding constraints coming from multiple production modes or final states. In sec.~\ref{sec:apps}, we apply these methods to experimental searches for spin-0, spin-1, and spin-2 resonances. Section~\ref{sec:simplex} discusses the use of N-simplex diagrams as a generalization of our method to cases where more than three branching ratios are sources of experimental constraints on the properties of new resonances. We conclude in sec.~\ref{sec:disc} and highlight the value of using the digital data record to enable exploration of an arbirtrary-dimensional parameter space of branching ratios.

  \section{Simplified Limits on Narrow Resonances}
  \label{sec:simpLimits}
  
  In this section, we review the formulation, presented in Refs.~\cite{Chivukula:2016hvp,Chivukula:2017lyk}, leading to the model-independent simplified parameter $\zeta$ allowing one to compare bounds from data with the easily calculated product of BRs corresponding to production and decay of a narrow resonance as well as its width-to-mass ratio. We begin with the resonance's partonic cross section.
  
  The tree-level cross section for resonant production of a state $R$ from initial state partons $i,\, j$ and decaying to final state $x,\, y$ can be written in the Breit-Wigner form as
  \begin{align}
    \hat{\sigma}_{ij \rightarrow R \rightarrow xy} (\hat{s}) = 16 \pi \mathcal{N}_{ij}  \left ( 1 + \delta_{ij} \right ) \frac{\Gamma_{xy} \, \Gamma_{ij}}{(\hat{s} - m_R^2)^2 + m_R^2 \Gamma_R^2} \, ,
  \end{align}
  with $\Gamma_{ab} \equiv \Gamma(R \rightarrow a b)$ the resonance's partial decay width, $\Gamma_R$ its total width, and $m_R$ its mass. Here, $\hat{s}$ is the partonic center of mass energy of the system, while $\mathcal{N}_{ij}$ is a ratio of spin and color factors,
  \begin{align}
    \mathcal{N}_{ij} \equiv \frac{N_{S_R}}{N_{S_i} N_{S_j}} \, \frac{C_R}{C_i C_j} \, ,
  \end{align}
  with $N_S$ and $C$, respectively, counting the number of spin and color states of the incoming partons and resonance. In the narrow width approximation (NWA), $\Gamma_R / m_R \ll 1$ and
  \begin{align}
    \frac{1}{(\hat{s} - m_R^2)^2 + m_R^2 \Gamma_R^2} \approx \frac{\pi}{m_R \Gamma_R} \delta (\hat{s} - m_R^2) \, .
  \end{align}
  Thus, the tree-level cross section in the NWA is given by
  \begin{align}
    \hat{\sigma}_{ij \rightarrow R \rightarrow xy} (\hat{s}) = 16 \pi^2 \mathcal{N}_{ij}  \left ( 1 + \delta_{ij} \right ) \textrm{BR}_{xy} \, \textrm{BR}_{ij} \, \frac{\Gamma_R}{m_R} \, \delta(\hat{s} - m_R^2) \, ,
  \end{align}
  with $\textrm{BR}_{ab} \equiv \Gamma_{ab} / \Gamma_R$ the BR of the resonance.
  
  For hadron colliders, the partonic cross section is related to the experimentally observable cross section by convolving it with the hadrons' parton distribution functions (PDFs). For proton-proton colliders like the LHC, we have
  \begin{align}
    \sigma_{p p \rightarrow R \rightarrow xy}(s) = 16 \pi^2 \mathcal{N}_{ij} \left ( 1 + \delta_{ij} \right ) \textrm{BR}_{ij} \, \textrm{BR}_{xy} \, \frac{\Gamma_R}{m_R} \left [ \frac{1}{s} \frac{d L_{ij}}{d \tau} \right ]_{\tau = \frac{m_R^2}{s}} \, ,
  \end{align}
  with $s$ the proton-proton center of mass energy. Here, $dL_{ij} / d\tau$ is the parton luminosity function,
  \begin{align}
    \frac{d L_{ij}}{d \tau} \equiv \frac{1}{1 + \delta_{ij}} \int_\tau^1 \frac{dx}{x} \left [ f_i(x,\, \mu_F^2) f_j \left (\frac{\tau}{x},\, \mu_F^2 \right ) + f_j(x,\, \mu_F^2) f_i \left (\frac{\tau}{x},\, \mu_F^2 \right ) \right ] \, ,
  \end{align}
  with $f_i$ the PDF for parton $i$, $x$ the fraction of the proton's momentum carried by the parton, and $\mu_F$ the factorization scale. If multiple partons contribute to the same experimental signal (e.g. light quark production or decay), this can be extended to a sum over initial and/or final state partons,
  \begin{align}
    \sigma_{pp \rightarrow R \rightarrow XY}(s) = 16 \pi^2 \sum_{i^\prime j^\prime} \textrm{BR}_{i^\prime j^\prime} \sum_{xy \in XY} \textrm{BR}_{xy} \, \frac{\Gamma_R}{m_R} \left [ \sum_{ij} \omega_{ij} \mathcal{N}_{ij} \frac{1 + \delta_{ij}}{s} \frac{dL_{ij}}{d \tau} \right ]_{\tau = \frac{m_R^2}{s}} \, ,
  \end{align}
  with $X Y$ the observable final state. The weight function $\omega_{ij}$,
  \begin{align}
    \omega_{ij} \equiv \frac{\textrm{BR}_{ij}}{\sum_{i^\prime j^\prime} \textrm{BR}_{i^\prime j^\prime}} \, ,
  \end{align}
  lies between 0 and 1 such that $\sum \omega_{ij} = 1$ by construction; $\omega_{ij}$ represents the fraction of the resonance's total production rate due to each individual partonic channel.
  
  We are now ready to define the $\zeta$ parameter,\footnote{The arrangement of kroenecker deltas in this definition of the $\zeta$ parameter differs slightly from that of Refs.~\cite{Chivukula:2016hvp,Chivukula:2017lyk}. This definition of $\zeta$ is more convenient because it has two distinct upper limits instead of four, depending only on whether or not the production modes also contribute to the observed final state.}
  \begin{align}\label{eq:zeta}
    \zeta \equiv \sum_{ij} \textrm{BR}_{ij} \sum_{xy \in XY} \textrm{BR}_{xy} \, \frac{\Gamma_R}{m_R} = \frac{\sigma_{p p \rightarrow R \rightarrow XY}(s)}{16 \pi^2} \left [ \sum_{ij} \omega_{ij} \mathcal{N}_{ij} \frac{1 + \delta_{ij}}{s} \frac{dL_{ij}}{d \tau} \right ]^{-1}_{\tau = \frac{m_R^2}{s}} \, .
  \end{align}
  Note that $\zeta$ retains much of the model-independence of the experimental search, which depends predominantly on the spin and helicity of the resonance, while translating the constraint into purely partonic quantities. As defined in eq.~\eqref{eq:zeta}, $\zeta$ is bounded from above by $\Gamma_R / m_R$ ($\Gamma_R / 4 m_R$) when there is overlap (no overlap) between initial and final states. As we are working in the NWA where $\Gamma_R / m_R \lesssim 10 \%$, this corresponds to an upper bound on $\zeta$ of $1/10$ ($1/40$) when there is overlap (no overlap) between initial and final states.

  \section{Ternary Diagrams}
  \label{sec:ternary}
  
  In this section, we introduce a prescription for extending the simplified limits framework to situations which necessitate considering constraints on more than a one dimensional parameter. We begin by discussing the situations considered in this paper: the limitations of the original simplified limits formulation and combining constraints from multiple independent experimental final states.

  The simplified parameter $\zeta$ defined in eq.~\ref{eq:zeta} offers a relatively model-independent conversion from the limits on $\sigma_\textrm{prod} \times \textrm{BR}$ to partonic quantities which more directly describe the properties of the resonance: its mass, width, and BRs. There are, however, several situations that strain the applicability of these one-dimensional limits, which we will explore here. The first exception is in situations where there may be more than one production mode. In these situations, deconvolving the hadronic PDFs introduces an ambiguity in constraints due to the weight factors $\omega_{ij}$, which are unknown \textit{a priori} without the introduction of additional model-dependent assumptions. An example of this would be Drell-Yan (DY) production of a heavy $Z^\prime$, which leads to a band in limits on $\zeta$. See the right panel of fig.~\ref{fig:qqWprimeZprimeZeta} in sec.~\ref{sec:spin1} for an example of this scenario.
  
  The second scenario we consider involves combining searches from multiple experimentally distinguishable final states. Combining the results of multiple searches offers increased statistics and is particularly effective when the experiments in question are similarly sensitive to two or more distinguishable final states. While monochannel searches place constraints on $\sigma_\textrm{prod} \times \textrm{BR}$, doing so for combined searches requires making an additional model-dependent assumption about the relationship between the various channels' BRs. The question we therefore wish to address is how best to represent model-independent constraints within the multidimensional space of parameters necessitated by the above two situations.
  
  In the framework of simplified limits, the relevant model-independent quantities are the resonance's BRs, its mass, and its total width. In combining channels, a model-independent scenario would be one where constraints on the properties of the resonance could be displayed within the space of BRs. Consider first the scenario of a resonance with one production mode and two decay modes. Here, we are naively forced to place limits within a three-dimensional space of BRs. However, we also have the sum rule
  \begin{align}
    \sum_{i = 1}^3 \textrm{BR}_i = 1 \, ,
  \end{align}
  reducing the space of independent BRs by one. Ternary diagrams,\footnote{Perhaps the most familiar example of a ternary diagram in particle physics is the Dalitz plot for three-body decays~\cite{Dalitz:1953cp}, where the sum of the two-body invariant masses of the final state is constrained by the kinematics of the system, e.g. for particle 0 decaying into particles 1, 2, and 3, the constraint is $m_{12}^2 + m_{13}^2 + m_{23}^2 = \sum_{i = 0}^3 m_i^2$.} representations of the space of three variables which sum to a constant, are ideally suited to displaying constraints within this parameter space.
  
\begin{figure}
    \centering
    \includegraphics[width=0.65\textwidth]{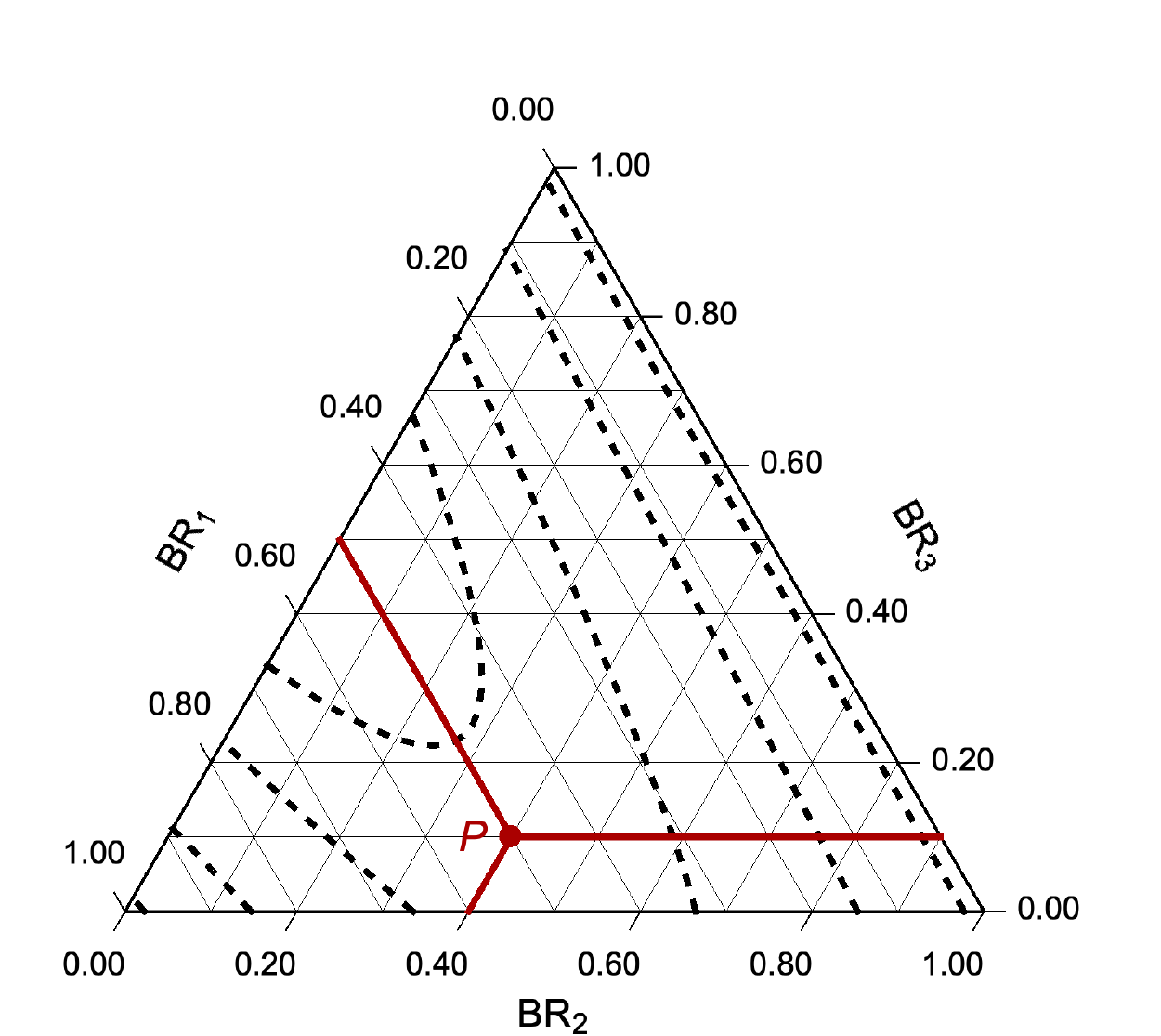}
    \caption{Example of a ternary diagram representing the space of BRs, with each point within the diagram satisfying the constraint $\sum_{i=1}^3 \textrm{BR}_i = 1$. The dashed lines represent hypothetical experimental constraints from combining limits, showing contours of constant $m_R$ or $\Gamma_R / m_R$. The point $P$ corresponds to $\{ \textrm{BR}_1,\, \textrm{BR}_2,\, \textrm{BR}_3 \} = \{0.5,\, 0.4,\, 0.1\}$.}
    \label{fig:ternaryEx1}
  \end{figure}

  Fig.~\ref{fig:ternaryEx1} shows an example of a ternary diagram, where at each point in the diagram the sum of BRs is one. The tick marks on the axes are skewed to indicate the lines of constant $\textrm{BR}_i$. Lines of constant $\textrm{BR}_2$ run parallel to the $\textrm{BR}_1$ axis, lines of constant $\textrm{BR}_3$ are parallel to the $\textrm{BR}_2$ axis, and lines of constant $\textrm{BR}_1$ are parallel to the $\textrm{BR}_3$ axis. The point $P$ is labeled as an example, corresponding to $\{ \textrm{BR}_1,\, \textrm{BR}_2,\, \textrm{BR}_3 \} = \{0.5,\, 0.4,\, 0.1\}$ with $\sum_{i = 1}^3 \textrm{BR}_i = 1$ by construction. Within this parameter space, one may then plot contours of upper limits on the remaining model parameters. In the simplified limits framework, one may therefore use ternary diagrams to display constraints on $m_R$ for fixed values of $\Gamma_R / m_R$ or on $\Gamma_R / m_R$ for fixed values of $m_R$. We present examples of ternary diagrams, displaying upper limits on $\Gamma_R / m_R$, in sec.~\ref{sec:apps}.

  Above, we mentioned the scenario where one production mode and two decay modes saturates all possible BRs. Interpreting the ternary diagrams for models of resonances with more than three BRs turns out to be straightforward. For a resonance with $n$ BRs, we instead have $\sum_{i = 1}^3 \textrm{BR}_i = 1 - \sum_{i = 4}^n \textrm{BR}_i$. We can thus define ``effective'' BRs,
  \begin{align}
    \widetilde{\textrm{BR}}_i \equiv \textrm{BR}_i \left ( 1 - \sum_{j = 4}^n \textrm{BR}_j \right )^{-1} \, ,
  \end{align}
  which satisfy the unitary sum rule, $\sum_{i = 1}^3 \widetilde{\textrm{BR}}_i = 1$, implicit in the construction of ternary diagrams. To interpret constraints displayed on the ternary diagrams in the framework of simplified limits, one must then also define an effective width,
  \begin{align}
    \widetilde{\Gamma}_R \equiv \Gamma_R \left ( 1 - \sum_{j = 4}^n \textrm{BR}_j \right )^2 \, .
  \end{align}
  A ternary diagram with sides spanning the range $[0,\,1]$ in this context generically displays the space of effective BRs, where $\widetilde{\textrm{BR}} \ge \textrm{BR}$ and $\widetilde{\Gamma}_R \le \Gamma_R$ with the equalities saturated only when the resonance has exactly three decay modes. In many cases it is expected that only a few channels will have similar experimental sensitivity over a wide range of the parameter space, and ternary diagrams then capture the most interesting region of the possible decay modes. However, a discussion of the generalization of this method to N-simplexes is presented in sec.~\ref{sec:simplex}.
  
  With this framework in place, we have introduced the means of extending searches for narrow resonances into the multidimensional parameter space of BRs. Within the simplified limits framework, this allows us to unfold the ambiguity introduced by deconvolving the hadronic PDFs when there is more than one production mode. In the case of multiple experimentally distinguishable final states, it allows us to combine search results without introducing further model-dependent assumptions about the relationship between BRs. In what follows, we explore specific examples of each of these situations.

  \section{Applications}
  \label{sec:apps}
  
  Searches for resonances in the diboson final state have been an important part of the search for BSM physics at the LHC, having been studied in detail by both ATLAS~\cite{Aad:2020tps,Aad:2020ddw,Aad:2019fbh} and CMS~\cite{Sirunyan:2019jbg}. Models usually considered for such searches include composite and little Higgs models~\cite{Marzocca:2012zn,Bellazzini:2014yua,Schmaltz:2005ky}, heavy vector triplet (HVT) models~\cite{deBlas:2012qp,Pappadopulo:2014qza}, and models of gravity in warped extra dimensions~\cite{Randall:1999ee,Agashe:2007zd}. The efficiency of a given search depends primarily on the spin and helicity of the resonance while remaining relatively model-independent otherwise. In this section, we explore the implications of searches for spin-0, 1, and 2 narrow resonances, using diboson final states as examples. We apply the simplified limits framework discussed above to existing searches by the ATLAS collaboration.
  
  For the conversion of experimental constraints and the calculation of the simplified limits parameter $\zeta$ we use the CT18 NLO central PDF set~\cite{Hou:2019efy}. The RS radion BM is calculated using leading-order formulae given in ref.~\cite{Barger:2011qn}. Possible large higher-order corrections to the radion's $gg$ coupling, parameterized by a K-factor, and heavy quark couplings are not included. Such corrections can be comparable to those of heavy Higgs production via gluon fusion where $\textrm{K} > 1$, leading to conservative estimates of the experimental constraints presented here. The HVT BMs are calculated at leading-order using CalcHEP 3.8.7~\cite{Belyaev:2012qa} with the model file provided by ref.~\cite{Pappadopulo:2014qza}. The RS graviton BM is calculated at leading-order using formulae given in refs.~\cite{Agashe:2007zd,Bijnens:2001gh}.
  
  Combining the statistics of multiple BSM physics searches requires delicate attention to the details of each experiment. In what follows, we do not attempt to reproduce a complete statistical analysis of the experiments considered. Instead, we make the conservative assumption that the combination of constraints from multiple channels is given by the strongest of the bounds,
  \begin{align}
    \sigma_\textrm{prod}^{95} = \textrm{min} \! \left [ \sigma_1^{95} , \, \sigma_2^{95} , \, \dots \right ] \, .
  \end{align}
  A detailed combination of the statistics of each search will in general produce stronger results, however such an analysis is beyond the scope of this paper. For more thorough discussions of the statistics involved in such searches, see e.g. refs.~\cite{Read:2002hq,Cowan:2010js}.

  \subsection{Spin-0 Resonance}
  \label{sec:spin0}
  
  We first consider a neutral spin-0 resonance ($\phi$) produced via gluon fusion. In ref.~\cite{Aad:2020ddw}, ATLAS reports constraints on the production of a Randall-Sundrum (RS) radion~\cite{Goldberger:1999uk} in the combined $WW+ZZ$ channels, with the ratio $\textrm{BR}(\phi \rightarrow WW) / \textrm{BR}(\phi \rightarrow ZZ)$ fixed by the model. The neutral scalar radion is a feature of extra-dimensional models which is predicted to stabilize the size of the extra dimension. The radion coupling to SM fields is inversely proportional to the vacuum expectation value of the radion field, $\Lambda_\phi$ and proportional to the mass (mass squared) of the SM fermions (bosons) it couples to. As the light fermion couplings to the radion are suppressed by their masses, the radion is predominantly produced via gluon fusion. We use $\Lambda_\phi = 3$~TeV and $k L = 35$ as a BM, where $k L$ is the size of the extra dimension. The radion's BRs are roughly constant for masses above a few TeV, with sizable BRs of
  \begin{align}
    \begin{array}{l l l}
    \textrm{BR}_{WW} = 47\% \, , \;\;\;\;\;\;\;\; &\textrm{BR}_{hh} = 23\% \, , \;\;\;\;\;\;\;\; & \textrm{BR}_{gg} = 5.8\% \, , \\
    \textrm{BR}_{ZZ} = 23\% \, , & \textrm{BR}_{t \overline{t}} = 0.9\% \, , & 
    \end{array}
  \end{align}
  at $M_\phi = 3$~TeV.

  \begin{figure}
    \centering
    \includegraphics[width=0.75\textwidth]{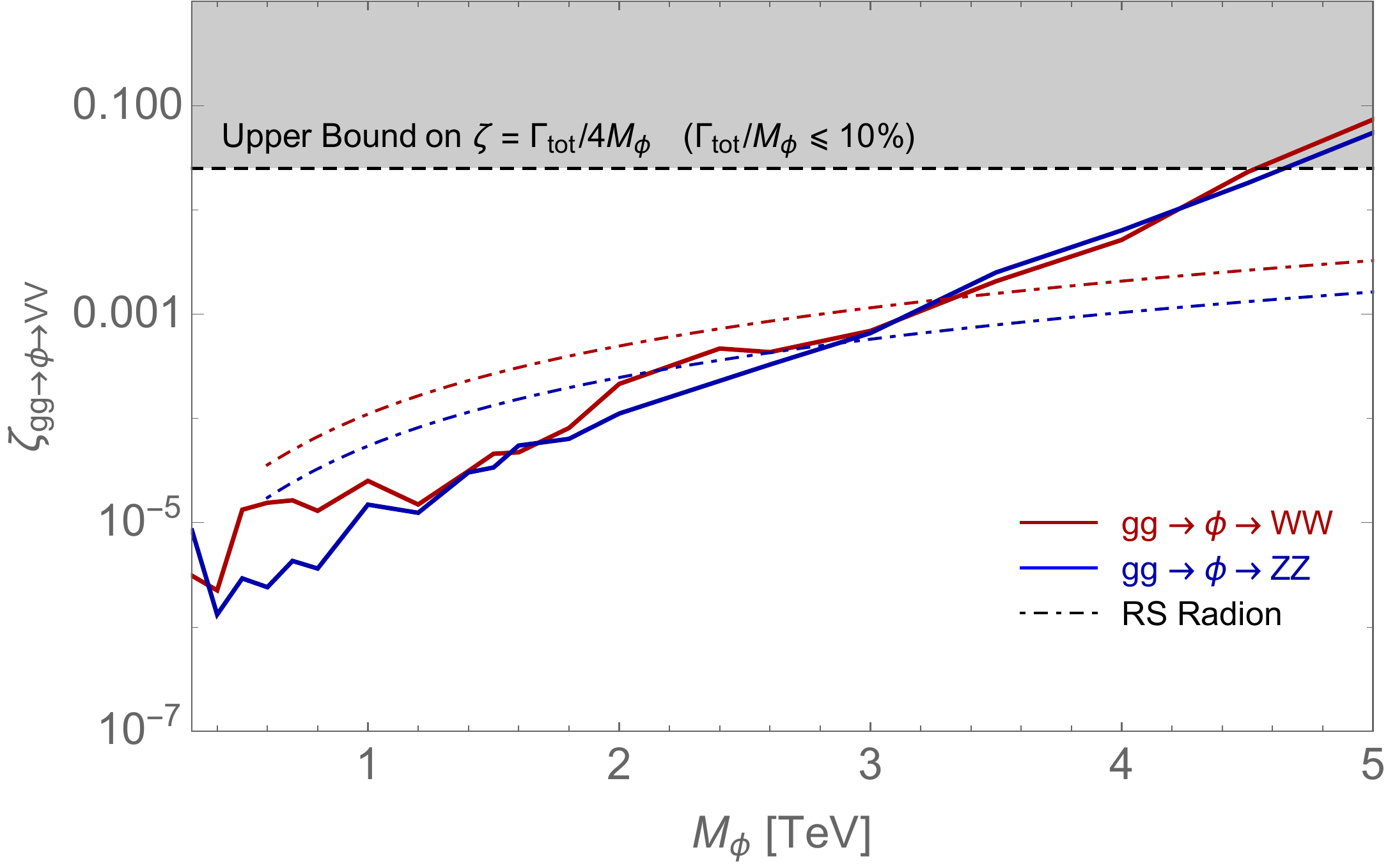}
    \caption{Constraints on narrow scalar resonance production from gluon fusion in the $WW$ and $ZZ$ channels~\cite{Aad:2020ddw}. Constraints from ATLAS are converted to upper limits on the $\zeta$ parameter, eq.~\eqref{eq:zeta}, represented by solid red and blue lines for the $WW$ and $ZZ$ channels, respectively. The gray shaded region corresponds to the upper limit on the product of branching ratios times $\Gamma_\phi / M_\phi = 10\%$, approximately where the NWA breaks down. The dot-dashed lines show the predictions for the radion in our BM model.}
    \label{fig:GGWWZZspin0zeta}
  \end{figure}

  Fig.~\ref{fig:GGWWZZspin0zeta} shows the ATLAS constraints from the individual channels converted into the language of simplified limits. Generally the constraints from each channel are quite competitive, except below $M_\phi \sim 1$~TeV and between approximately 2.0 -- 2.6~TeV where the $ZZ$ channel is significantly more constraining. Above $M_\phi \sim 4.5$~TeV, the search does not constrain any model which satisfies the NWA assumption. Also shown are the predictions for our BM radion, which sets limits on the mass of the radion of $M_\phi \gtrsim 3.3~(2.9)$~TeV in the $WW$ ($ZZ$) channel. Other models would be represented by different curves in $\zeta$ vs. $M_\phi$, corresponding to different limits on the mass of the resonance.

  \begin{figure}
    \centering
    \includegraphics[width=0.75\textwidth]{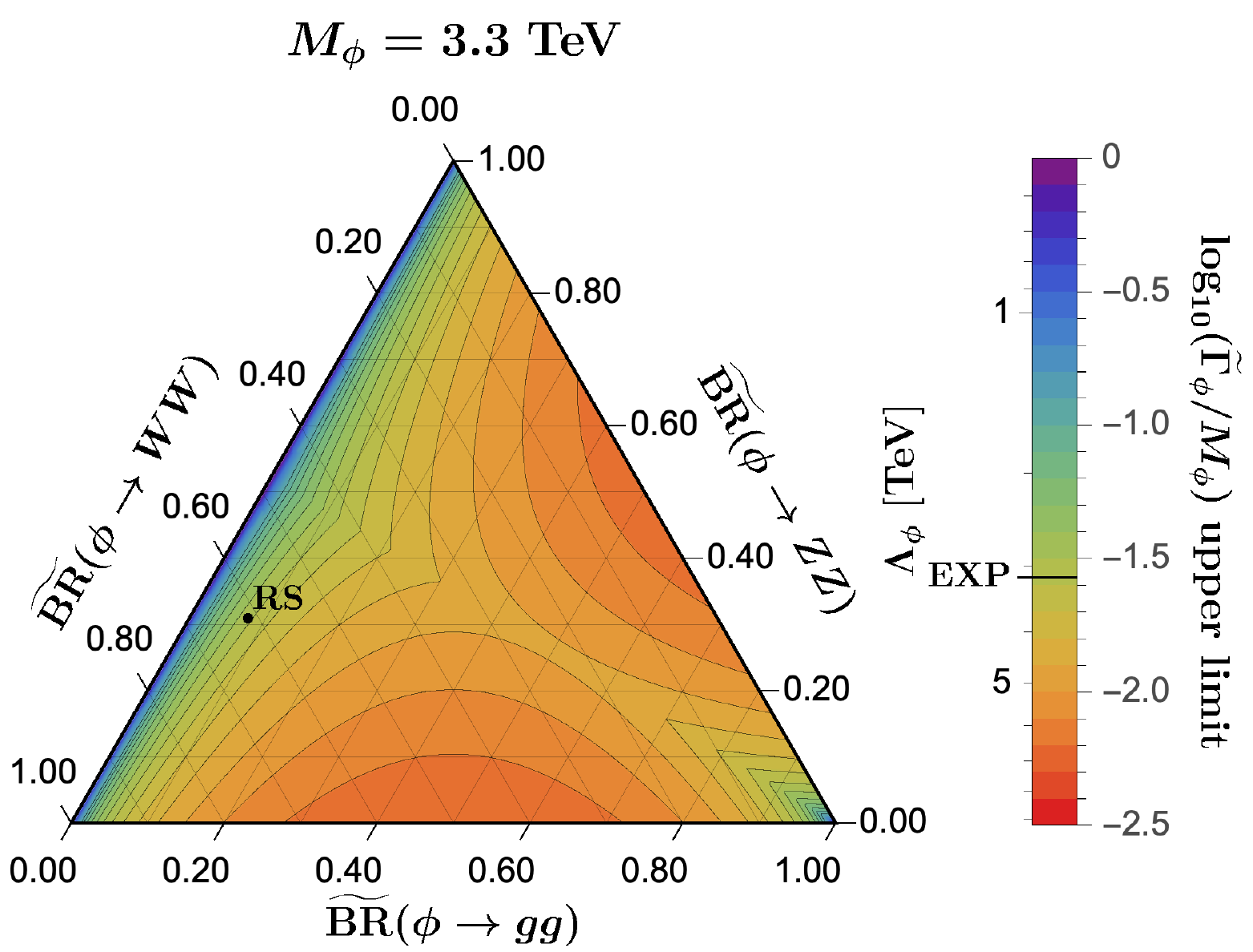}
    \caption{Ternary diagram showing constraints on a narrow scalar resonance produced via gluon fusion and decaying to $WW$ and $ZZ$~\cite{Aad:2020ddw} for a resonance mass of 3.3~TeV. The colors depict constraints on $\widetilde{\Gamma}_\phi / M_\phi$ (labeled right of the legend), which have been translated to the corresponding constraint on $\Lambda_\phi$ for the BM value $kL = 35$ (labeled left of the legend). The black dot represents the location in the parameter space of BRs corresponding to the radion in our BM model, and the black line through the legend labels the value of the experimental constraint on the BM radion. The BM prediction $\widetilde{\Gamma_\phi}/M_\phi \sim 10^{-1.5}$ (corresponding to $\Lambda_\phi=3$~TeV), being slightly above the the experimental limit, demonstrates that the model is just slightly excluded.}
    \label{fig:spin0tern}
  \end{figure}

  Applying the principles of sec.~\ref{sec:ternary},  fig.~\ref{fig:spin0tern} shows constraints on $\widetilde{\Gamma}_\phi / M_\phi$ for $M_\phi = 3.3$~TeV, which is near the experimental limit for our BM radion. The radion BM values are labeled by a point in the plane of $\widetilde{\textrm{BR}}$s and by a line on the legend of $\widetilde{\Gamma}_\phi / M_\phi$. In general, one can see that as the BRs for either the production mode or both decay modes decrease, the constraints unsurprisingly also weaken. Conversely, constraints are strongest when $\widetilde{\textrm{BR}}(\phi \rightarrow gg) \sim \widetilde{\textrm{BR}}(\phi \rightarrow VV)$ for a single channel, with the other channel's BR negligible. Contours of constant $\log_{10} \widetilde{\Gamma}_\phi / M_\phi$ are shown in increments of $10^{-1}$, and the BM value $\widetilde{\Gamma}_\phi / M_\phi \sim 10^{-1.5}$ roughly corresponds to the experimental limit, which is labeled by a solid black line through the legend. Sharp edges in the contours of constant $\widetilde{\Gamma}_\phi / M_\phi$ occur when the constraints from both channels are equal, and highlight the region of parameter space where one would expect to gain the most from a proper statistical combination of the independent searches. As the BM radion model also predicts a sizable $\textrm{BR}(\phi \rightarrow hh)$, we also see the application of effective BRs, where $\widetilde{\textrm{BR}}_{WW}\; (\textrm{BR}_{WW}) = 62\%\; (47\%)$, $\widetilde{\textrm{BR}}_{ZZ}\; (\textrm{BR}_{ZZ}) = 31\%\; (23\%)$, and $\widetilde{\textrm{BR}}_{gg}\; (\textrm{BR}_{gg}) = 7.7\%\; (5.8\%)$.

  \begin{figure}
    \includegraphics[width=0.475\textwidth]{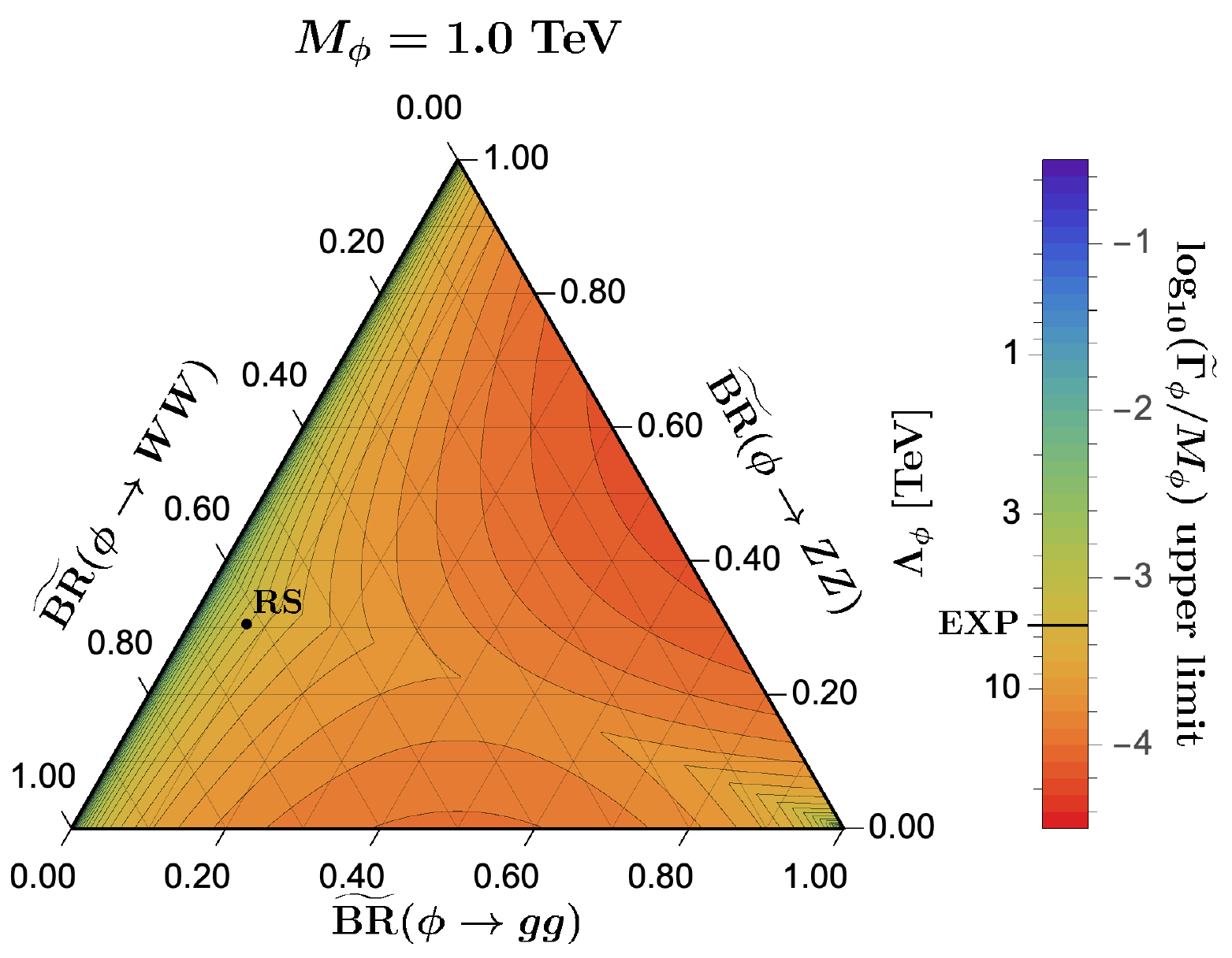}
    \hspace*{0.025\textwidth}
    \includegraphics[width=0.475\textwidth]{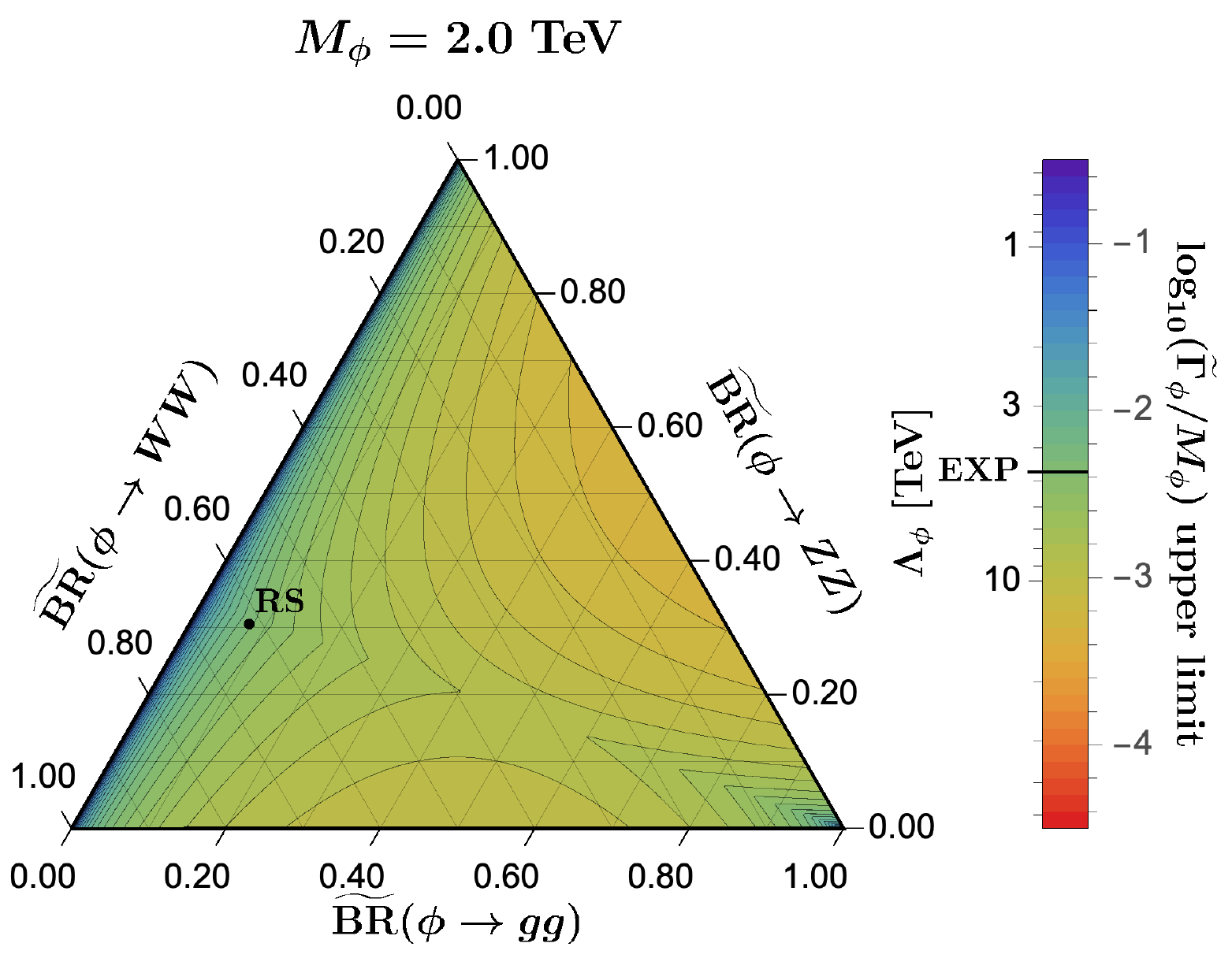}
    \vspace*{5mm} \\
    \includegraphics[width=0.475\textwidth]{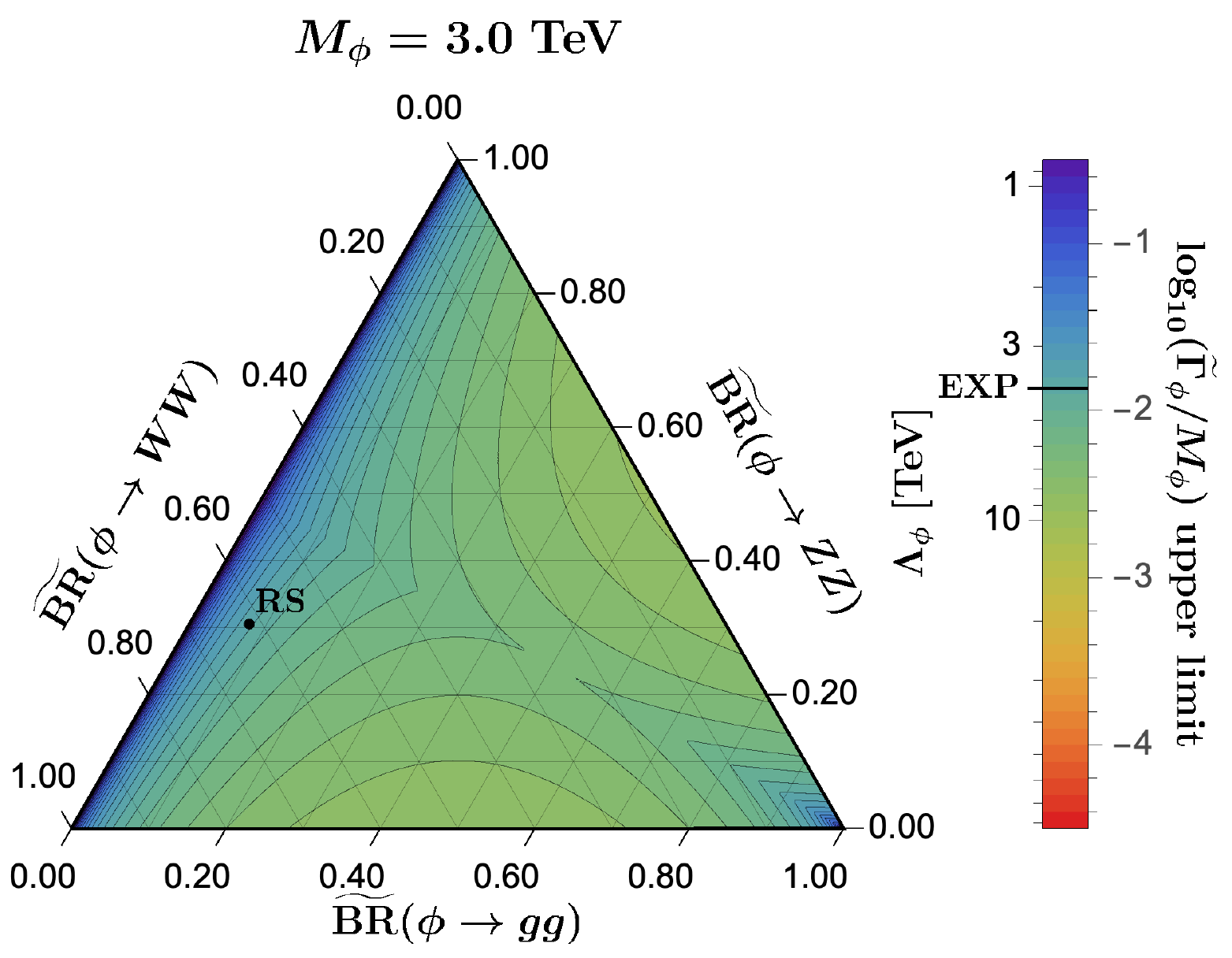}
    \hspace*{0.025\textwidth}
    \includegraphics[width=0.475\textwidth]{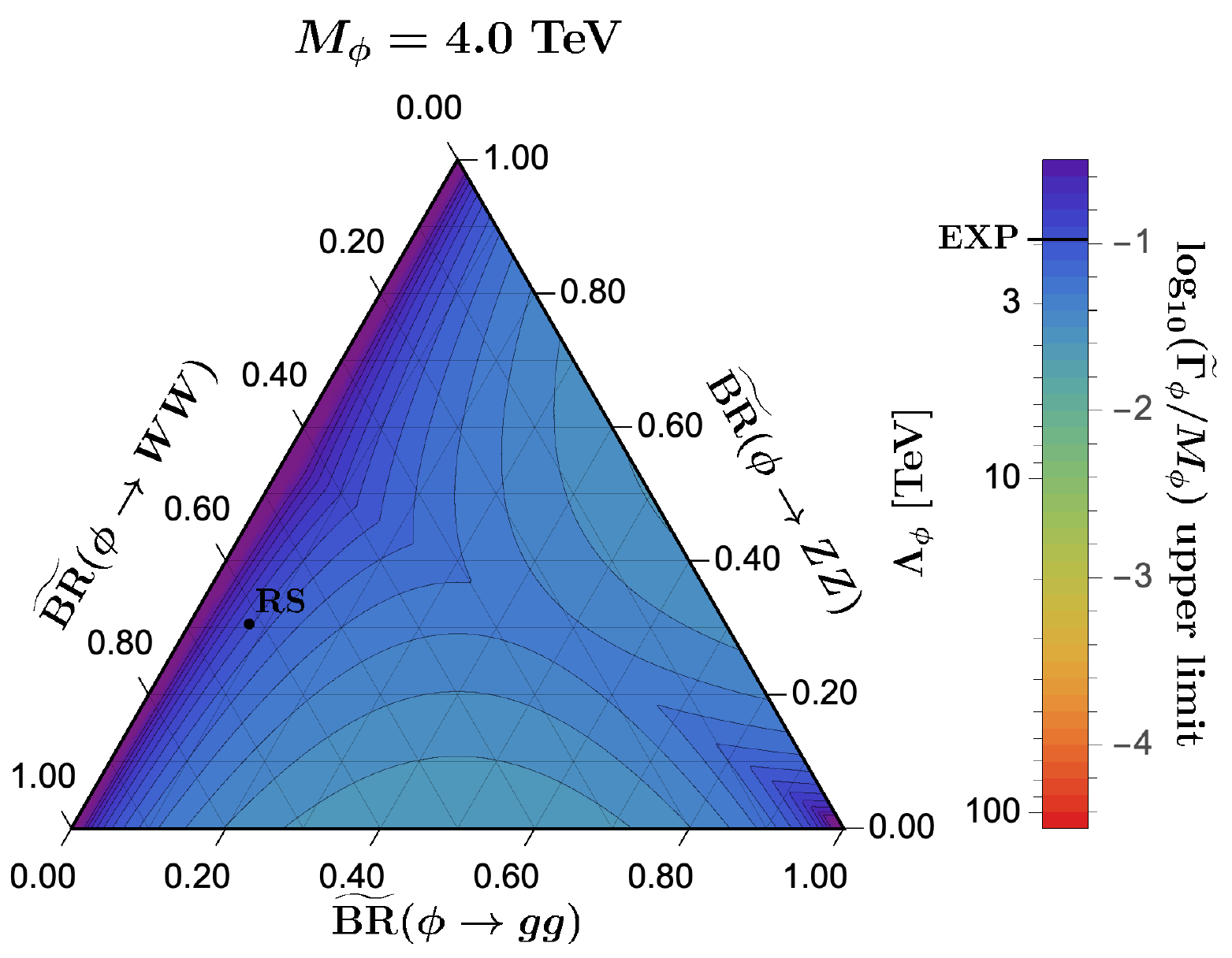}
    \caption{Ternary diagrams showing constraints on scalar narrow resonance production from gluon fusion in the $WW$ and $ZZ$ channels~\cite{Aad:2020ddw} for a selection of resonance masses. The colors depict constraints on $\widetilde{\Gamma}_\phi / M_\phi$ (labeled right of the legends), which have been translated to the corresponding constraint on $\Lambda_\phi$ for the BM value $kL = 35$ (labeled left of the legends). The black dot represents the location in the parameter space corresponding to the radion in our BM model, and the black line through the legend labels the value of the experimental constraint on the BM radion.}
    \label{fig:GGWWZZspin0ternary}
  \end{figure}

  To demonstrate how the diagrams change as a function of mass, fig.~\ref{fig:GGWWZZspin0ternary} shows constraints on $\widetilde{\Gamma}_\phi / M_\phi$ for several choices of resonance mass between $1.0~\textrm{TeV} \le M_\phi \le 4.0~\textrm{TeV}$. Here one can clearly see how the constraints weaken as the mass of the resonance increases, while within the space of $\widetilde{\textrm{BR}}$s the relative trends remain similar to those discussed above for fig.~\ref{fig:spin0tern}. Because the radion couplings are all proportional to $\Lambda_\phi^{-1}$, increasing $\Lambda_\phi$ decreases the radion's total width and vice versa, while keeping individual BRs constant at leading order. Therefore, each point on the ternary diagram also translates directly into a constraint on $\Lambda_\phi$ for a fixed value of $k L$. Constraints on $\Lambda_\phi$ for the BM value $k L = 35$ are also labeled to the left of the legend of each diagram, and the experimental constraint on the BM point is labeled by a solid black line through the legend. For the BM value $\Lambda_\phi = 3$~TeV, the radion is shown to be excluded in the diagrams corresponding to $M_\phi \le 3.0$~TeV, while it is unconstrained in the $M_\phi = 4.0$~TeV diagram.

  \subsection{Spin-1 Resonance}
  \label{sec:spin1}
  
  For a spin-1 resonance, we consider the production and decay of either a charged ($W^{\prime\,\pm}$) or neutral ($Z^\prime$) narrow vector resonance. Here gluon fusion production is forbidden by charge conservation or Yang's theorem, so production occurs via either Drell-Yan (DY) or vector boson fusion (VBF) processes. For DY production of a charged vector resonance, production occurs via $q \overline{q}^\prime$ annihilation and in most cases is dominated by the valence quark combination $u \overline{d}$ or $d \overline{u}$. For a neutral vector resonance, however, production occurs via $q \overline{q}$ and is dominated by a combination of valence quarks, $u \overline{u} + d \overline{d}$. As the ratio of couplings of the neutral resonance to up and down quarks is not known \textit{a priori}, deconvolving the proton PDFs introduces an ambiguity to the $\zeta$ parameter, as discussed in sec.~\ref{sec:ternary}.
  
  The production and decay of $W^{\prime\,\pm}$ and $Z^\prime$ can be parameterized in terms of an HVT model, which is a phenomenological framework proposed by the authors of ref.~\cite{Pappadopulo:2014qza} to cover a variety of explicit BSM models. In the HVT model, an $SU(2)_L$ vector triplet $V^\prime$ is introduced, and its interactions with the SM fields are parameterized by a variety of couplings. The parameter $g_V$ characterizes the typical interaction strength of the new triplet, and the parameters $c_H$ and $c_F$ characterize deviations from this strength in coupling to the Higgs and fermion currents, respectively. A factor of $g^2 / g_V^2$ is inserted in the coupling of $V^\prime$ to the SM fermions to make contact with many specific extended gauge models in the literature, where $g$ denotes the $SU(2)_L$ gauge coupling. Therefore, the interaction of $V^\prime$ with the Higgs doublet current (and therefore with the longitudinal components of the SM $W$ and $Z$) is parameterized by $(c_H g_v)$. Likewise, the coupling of $V^\prime$ to the SM fermions is controlled by the combination $(g^2/g_V) c_f$. 
  
  BM values are typically chosen to represent the range of various specific BSM extensions. Model A, with $g_V = 1$, is representative of a weakly coupled scenario such as in theories with an extended gauge symmetry. Model B, with $g_V = 3$, is chosen to represent a strongly-coupled composite Higgs scenario.\footnote{Our BM values for $c_F$ and $c_H$ are calculated from the relations in appendix~A of ref.~\cite{Pappadopulo:2014qza} with $\widetilde{c}_{VW} = -1$, $\widetilde{c}_H = \widetilde{c}_F = 0$ for model A and $\widetilde{c}_{VW} = -\widetilde{c}_H = 1$, $\widetilde{c}_F = 0$ for model B.} In both BM scenarios we assume a universal coupling of $V^\prime$ to all SM fermions; all interactions not mentioned above, which only contribute to the decay of $V^\prime$ via the small mixing between $V^\prime$ and the SM weak gauge bosons, are turned off. The two BM models predict dramatically different decays for the heavy spin-one resonance. Model A predicts a BR of a few percent to bosons, with dijets representing the dominant decay mode, while model B predicts decays predominantly into dibosons.

  \begin{figure}
    \includegraphics[width=0.4875\textwidth]{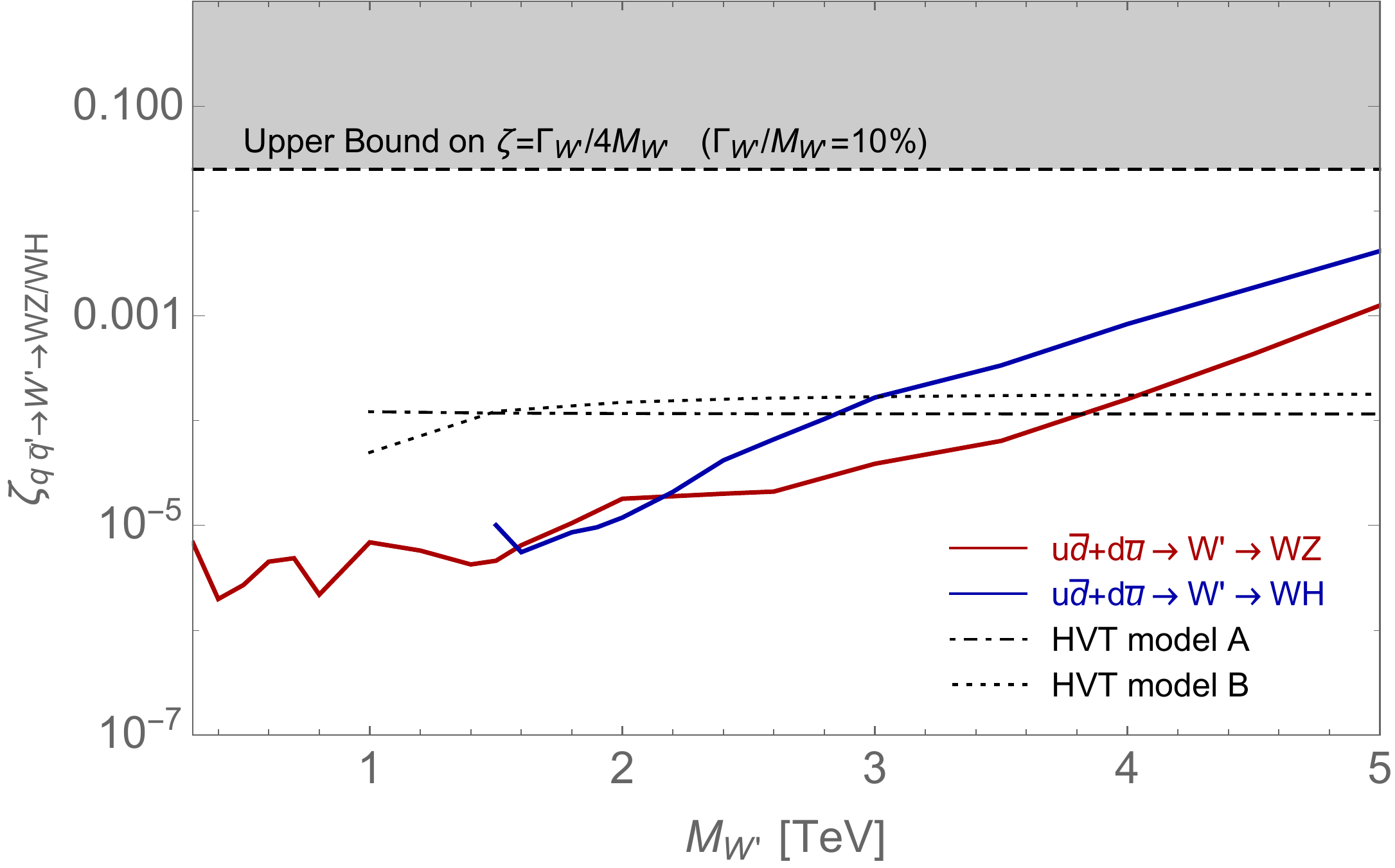}
    \hspace*{0.025\textwidth}
    \includegraphics[width=0.4875\textwidth]{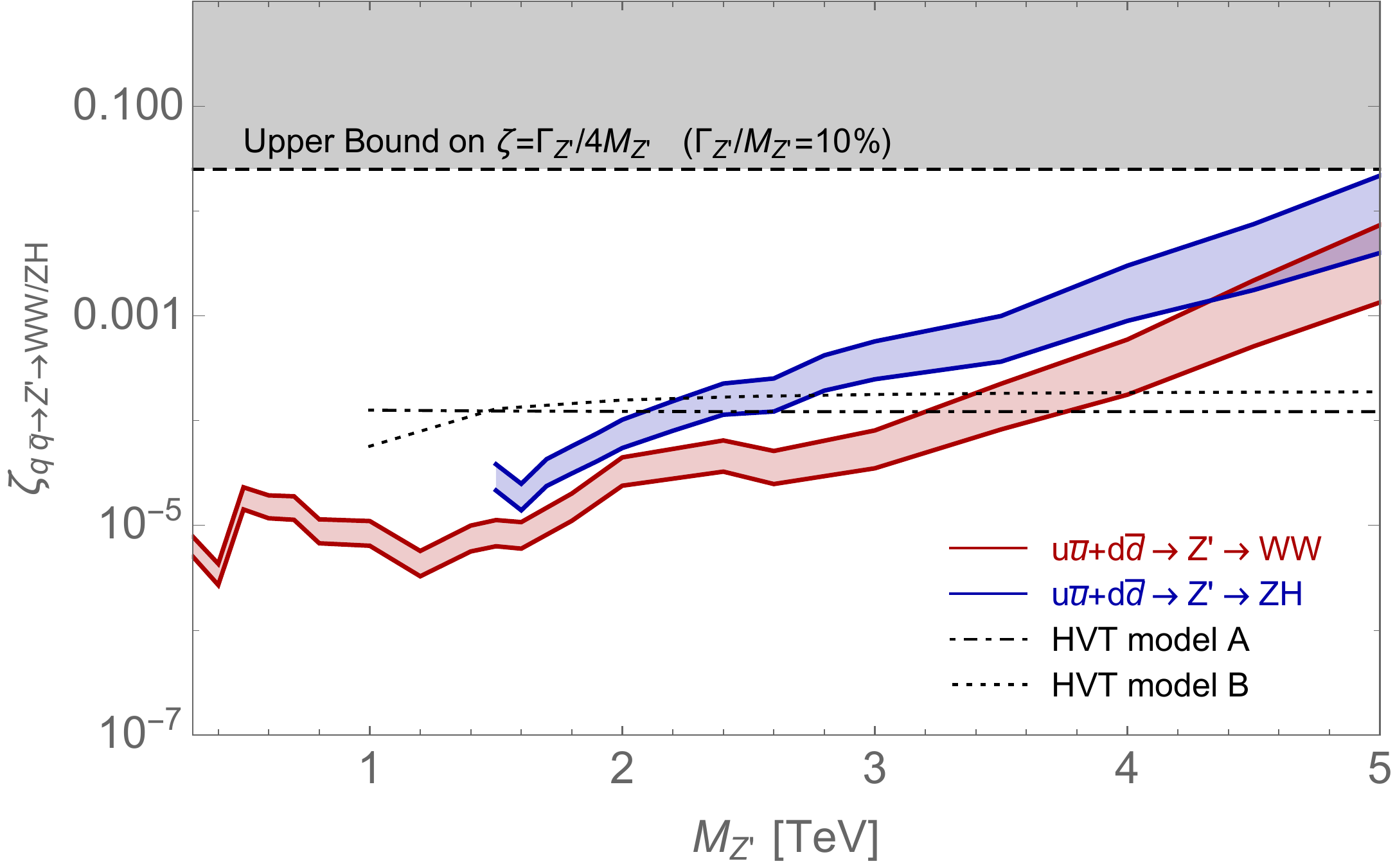}
    \caption{Constraints on charged (left pane) and neutral (right pane) narrow vector resonances from DY production in the $WV$~\cite{Aad:2020ddw} and $VH$~\cite{Aad:2020tps} channels. Constraints from ATLAS are converted to upper limits on the $\zeta$ parameter, eq.~\eqref{eq:zeta}, represented by solid red and blue lines for the $WV$ and $VH$ channels, respectively. The shaded bands in the constraints of the right plot are due to the model-dependent relationship between the production mode couplings of the $Z^\prime$, with the lower constraint corresponding to purely $u \overline{u}$ production and the upper constraint corresponding to purely $d \overline{d}$ production. The gray shaded region corresponds to the upper limit on the product of branching ratios times $\Gamma_{V^\prime} / M_{V^\prime} = 10\%$, where the NWA is no longer valid. Also shown are the HVT $p p \rightarrow V^\prime \rightarrow WV$ theory predictions for BM models A (black dot-dashed line) and B (black dotted line).}
    \label{fig:qqWprimeZprimeZeta}
  \end{figure}

  We recast constraints from ATLAS narrow resonance searches in the $WV$~\cite{Aad:2020ddw} and $VH$~\cite{Aad:2020tps} channels in terms of the simplified limits framework. The left plot of fig.~\ref{fig:qqWprimeZprimeZeta} shows constraints on the production of a narrow $W^\prime$ resonance in the $WZ$ and $WH$ channels, as well as the $W^\prime \rightarrow WZ$ predictions for the HVT BM models A and B. Although not shown, predictions for the $W^\prime \rightarrow WH$ channel are quite similar. We see that the $WZ$ channel dominates the constraints, with a narrow $W^\prime$ excluded for masses below 3.8 (4.0)~TeV for model A (model B). The right plot of fig.~\ref{fig:qqWprimeZprimeZeta} shows constraints on the production of a narrow $Z^\prime$ resonance in the $WW$ and $ZH$ channels, as well as the $Z^\prime \rightarrow WW$ predictions for the HVT BM models A and B. Here, the band in the constraints is due to the model-dependent relationship between the up and down quark couplings to $Z^\prime$. The lower limit of $\zeta$ corresponds to $\omega_{u \overline{u}} = 1$ and $\omega_{d \overline{d}} = 0$, while the upper limit of $\zeta$ corresponds to $\omega_{d \overline{d}} = 1$ and $\omega_{u \overline{u}} = 0$ with other values of $\omega$ falling between these two extremes. The HVT $Z^\prime$ is seen to be excluded for masses below $3.1~(3.3)~\textrm{TeV} \lesssim M_{Z^\prime} \lesssim 3.7~(4.0)~\textrm{TeV}$ for model A (model B).

  \begin{figure}
    \includegraphics[width=0.475\textwidth]{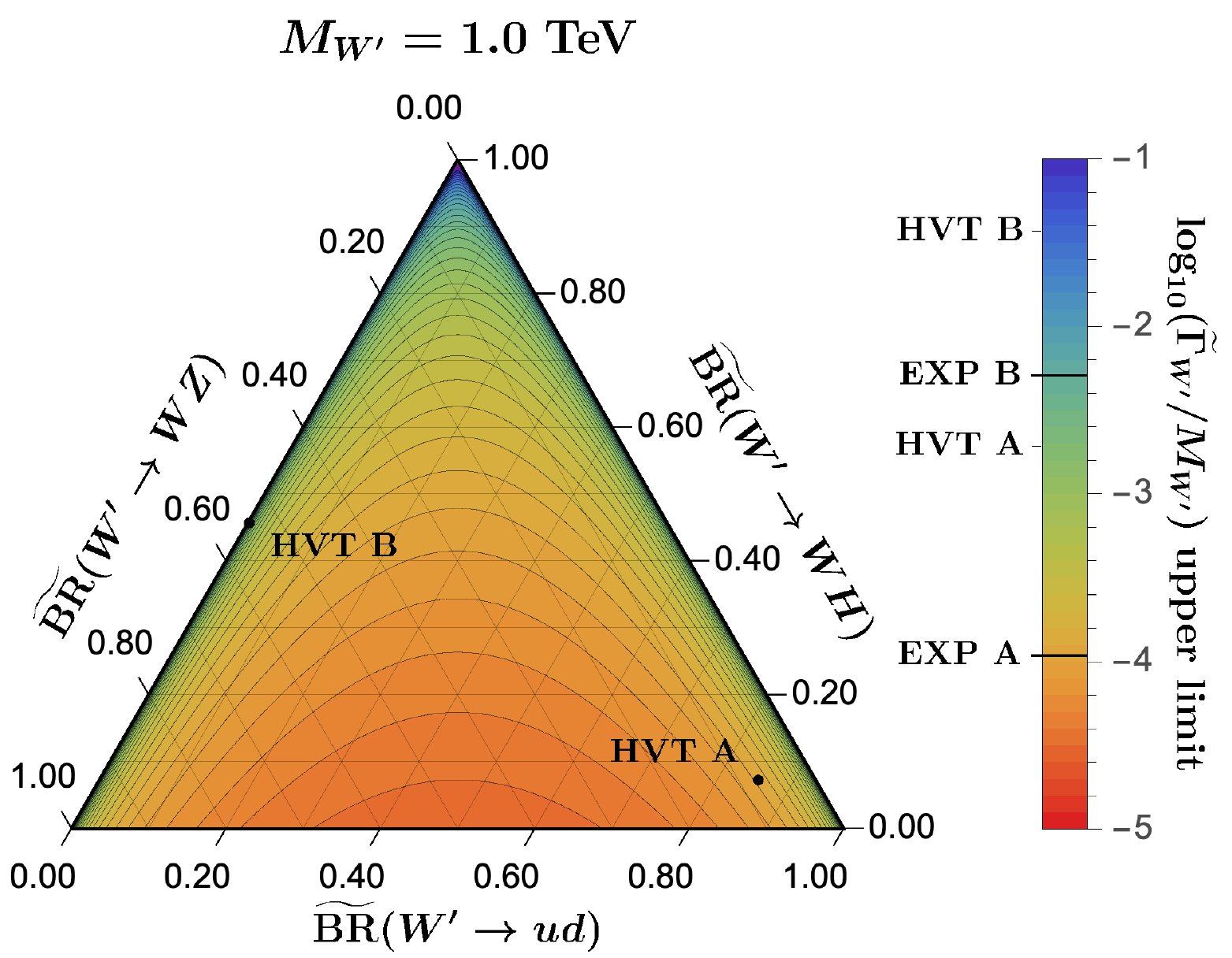}
    \hspace*{0.025\textwidth}
    \includegraphics[width=0.475\textwidth]{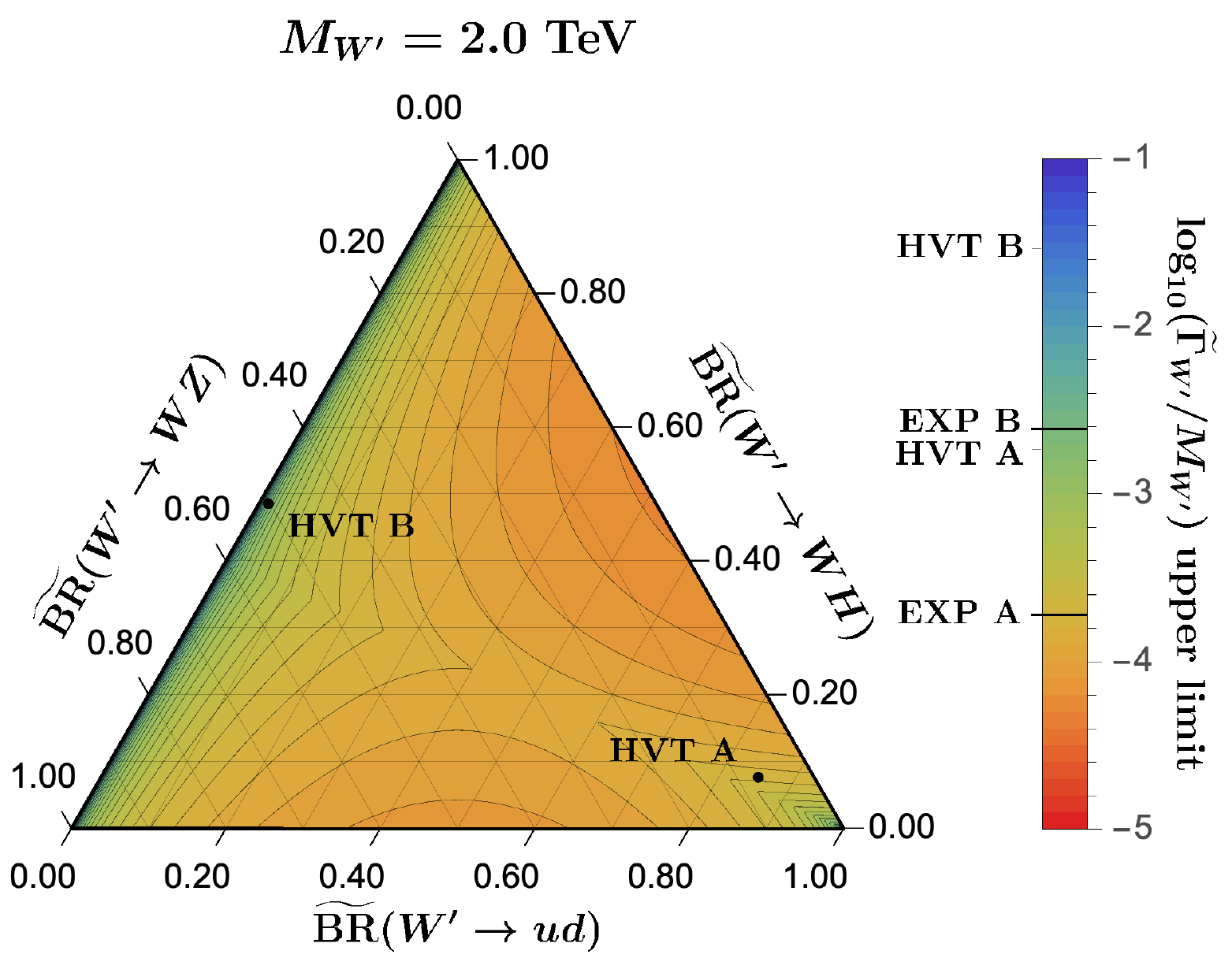}
    \vspace*{5mm} \\
    \includegraphics[width=0.475\textwidth]{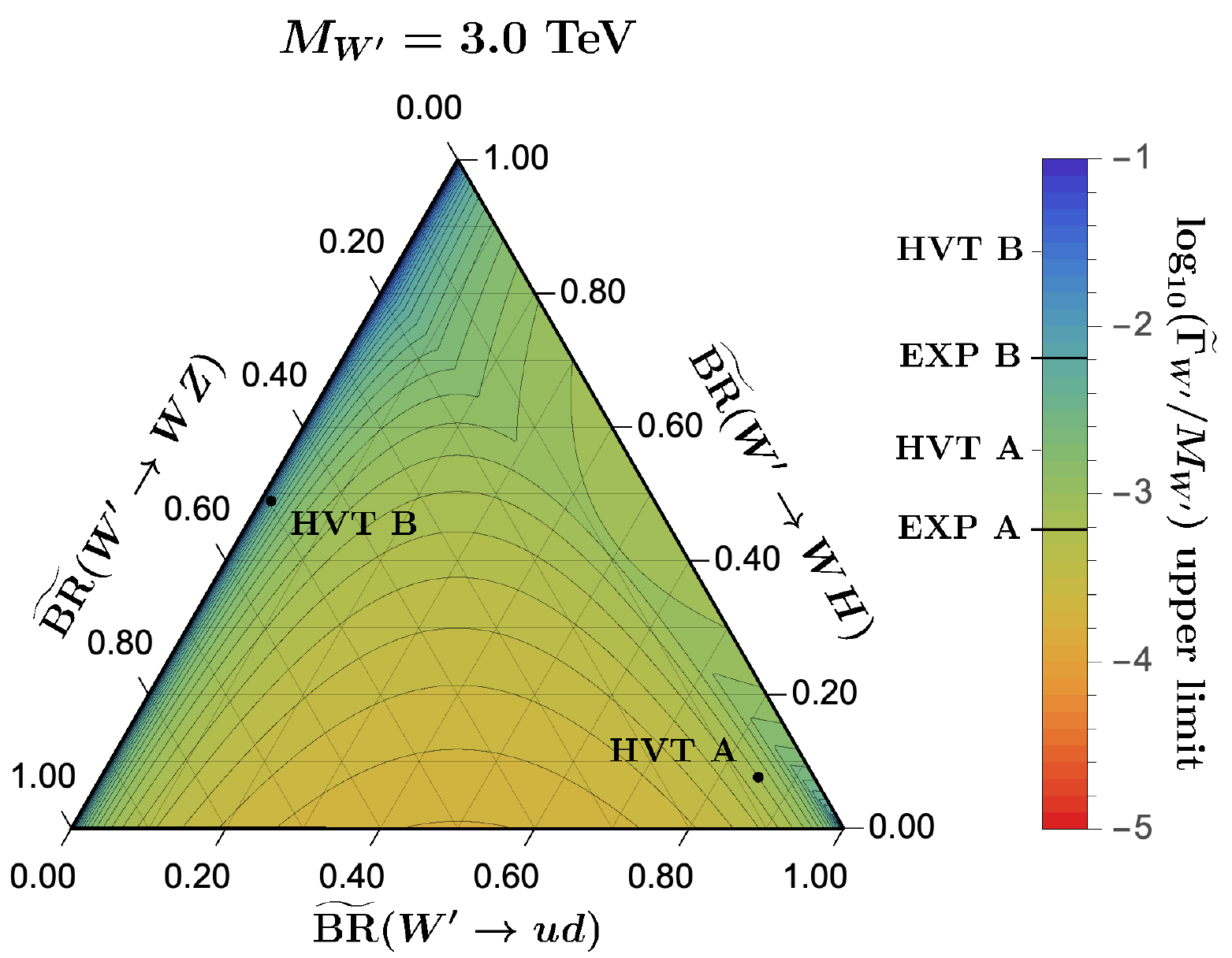}
    \hspace*{0.025\textwidth}
    \includegraphics[width=0.475\textwidth]{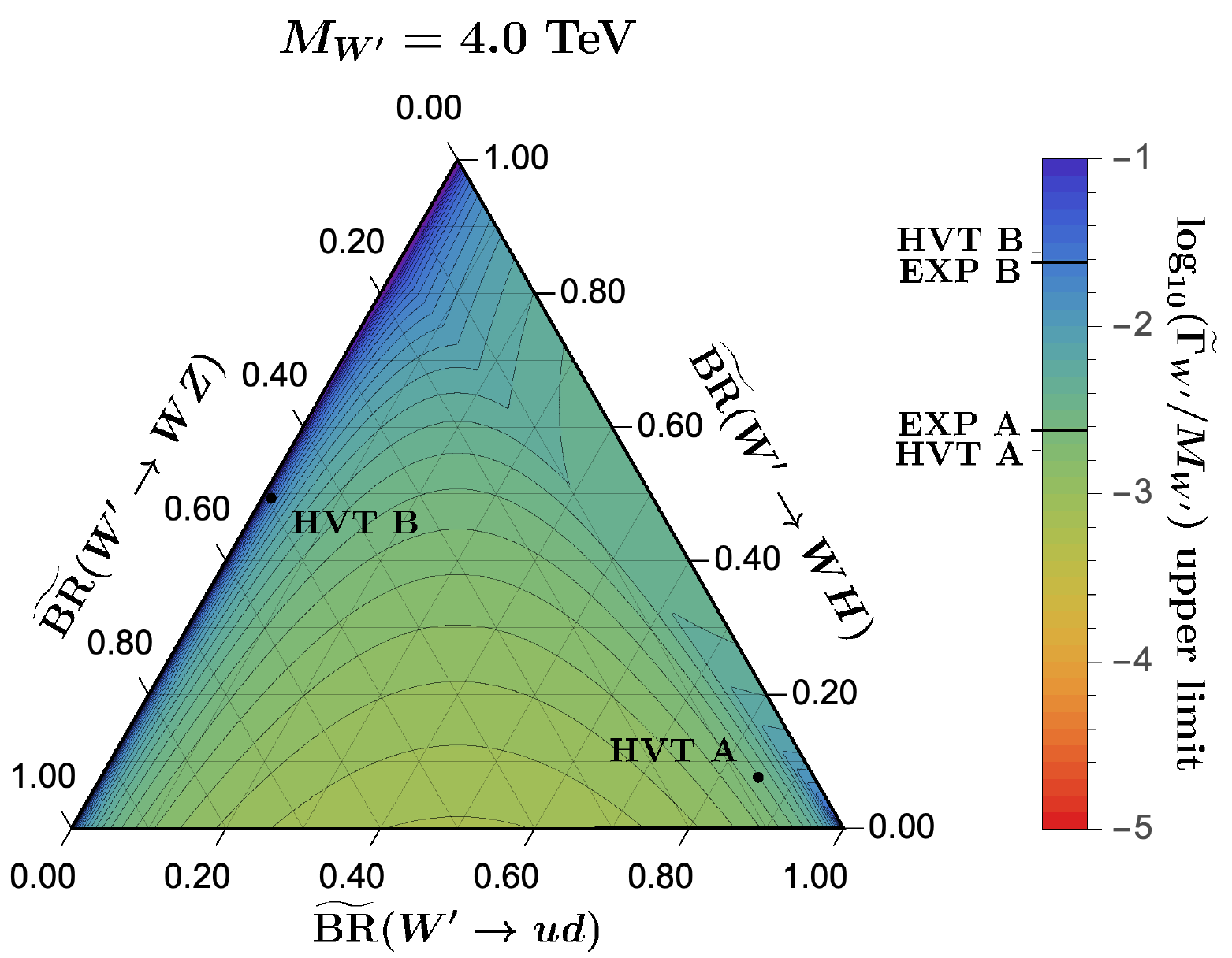}
    \caption{Ternary diagrams showing constraints on a charged vector narrow resonance from DY production in the $WZ$~\cite{Aad:2020ddw} and $WH$ channels~\cite{Aad:2020tps}. The colors depict constraints on $\widetilde{\Gamma}_{W^\prime} / m_{W^\prime}$ for a selection of resonance masses. Points labeled ``HVT A'' and ``HVT B'' mark the locations in the parameter space corresponding to the BMs for model A and model B, respectively. The black lines through the legend (labeled ``EXP A'' and ``EXP B'') show the values of the experimental constraints on the BM models, while the tick marks to the left of the legend (labeled ``HVT A'' and ``HVT B'') show their predicted values.}
    \label{fig:qqWprimeSpin1ternary}
  \end{figure}

  As the production of a charged vector resonance suffers no ambiguity in the initial state, we may apply the same principles used in sec.~\ref{sec:spin0} to combine limits from the $WZ$ and $WH$ channels. Fig.~\ref{fig:qqWprimeSpin1ternary} shows the combined constraints on $\widetilde{\Gamma}_{W^\prime} / M_{W^\prime}$ for several choices of resonance mass. At $M_{W^\prime} = 1.0$~TeV, there are no constraints from the $WH$ search, so the ternary diagram displays only constraints from the $WZ$ channel. On the other hand, the constraints at $M_{W^\prime} = 2.0$~TeV are similar for both channels which is reflected in the diagram. The remaining diagrams are again dominated by the $WZ$ channel, in agreement with the left panel of fig.~\ref{fig:qqWprimeZprimeZeta}. The HVT BMs are also displayed on the ternary diagrams, both by points labeling their predicted branching ratios and by tick marks on the left of legend labeling the predicted values of $\log_{10} ( \widetilde{\Gamma}_{W^\prime} / M_{W^\prime} )$. The solid lines through the legends label the experimental constraints on the BM points, so that one can more easily see that both BMs are excluded for $M_{W^\prime} \le 3.0$~TeV, while model A is unconstrained at $M_{W^\prime} = 4.0$~TeV and model B is very close to the experimental constraint.

  \begin{figure}
    \includegraphics[width=0.475\textwidth]{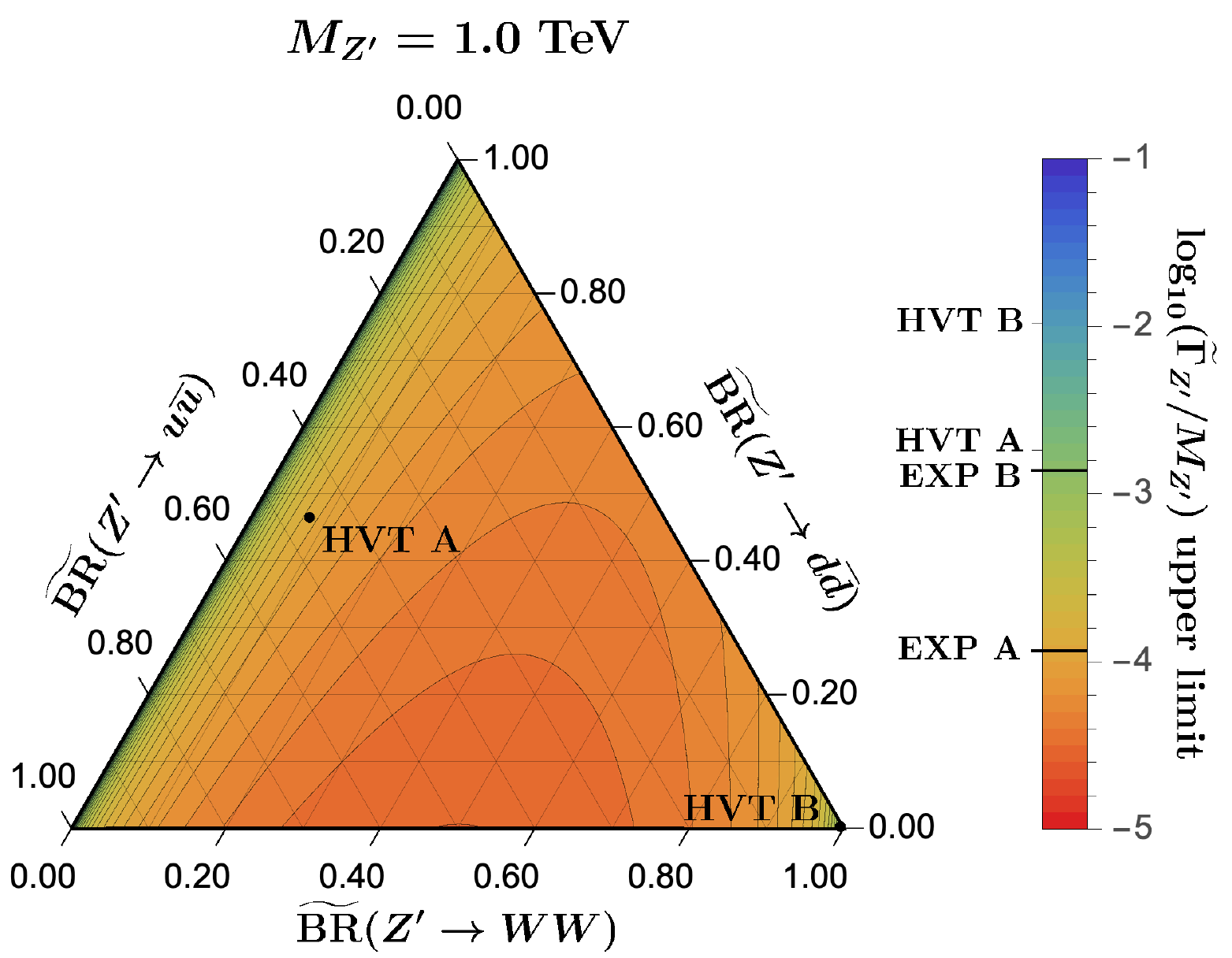}
    \hspace*{0.025\textwidth}
    \includegraphics[width=0.475\textwidth]{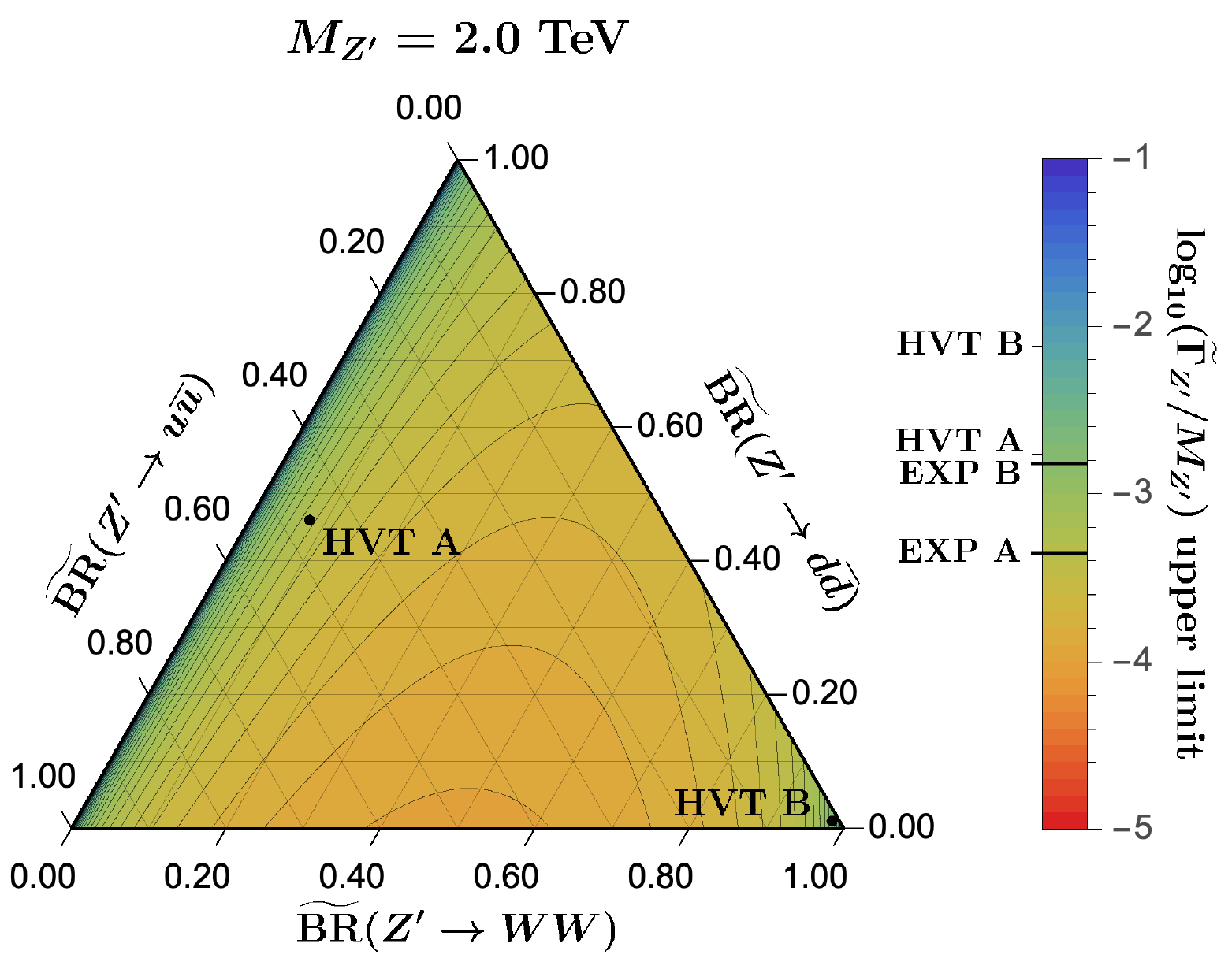}
    \vspace*{5mm} \\
    \includegraphics[width=0.475\textwidth]{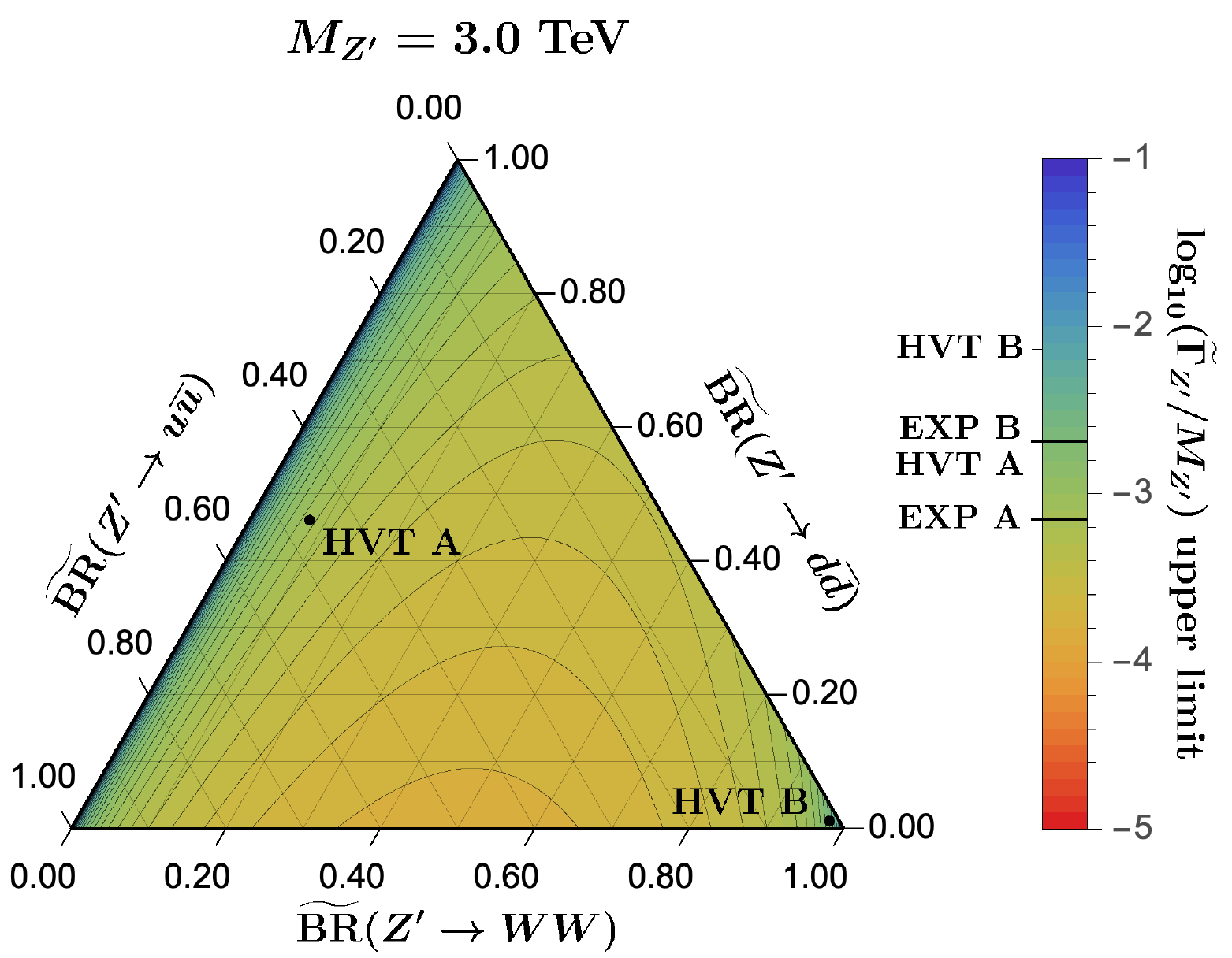}
    \hspace*{0.025\textwidth}
    \includegraphics[width=0.475\textwidth]{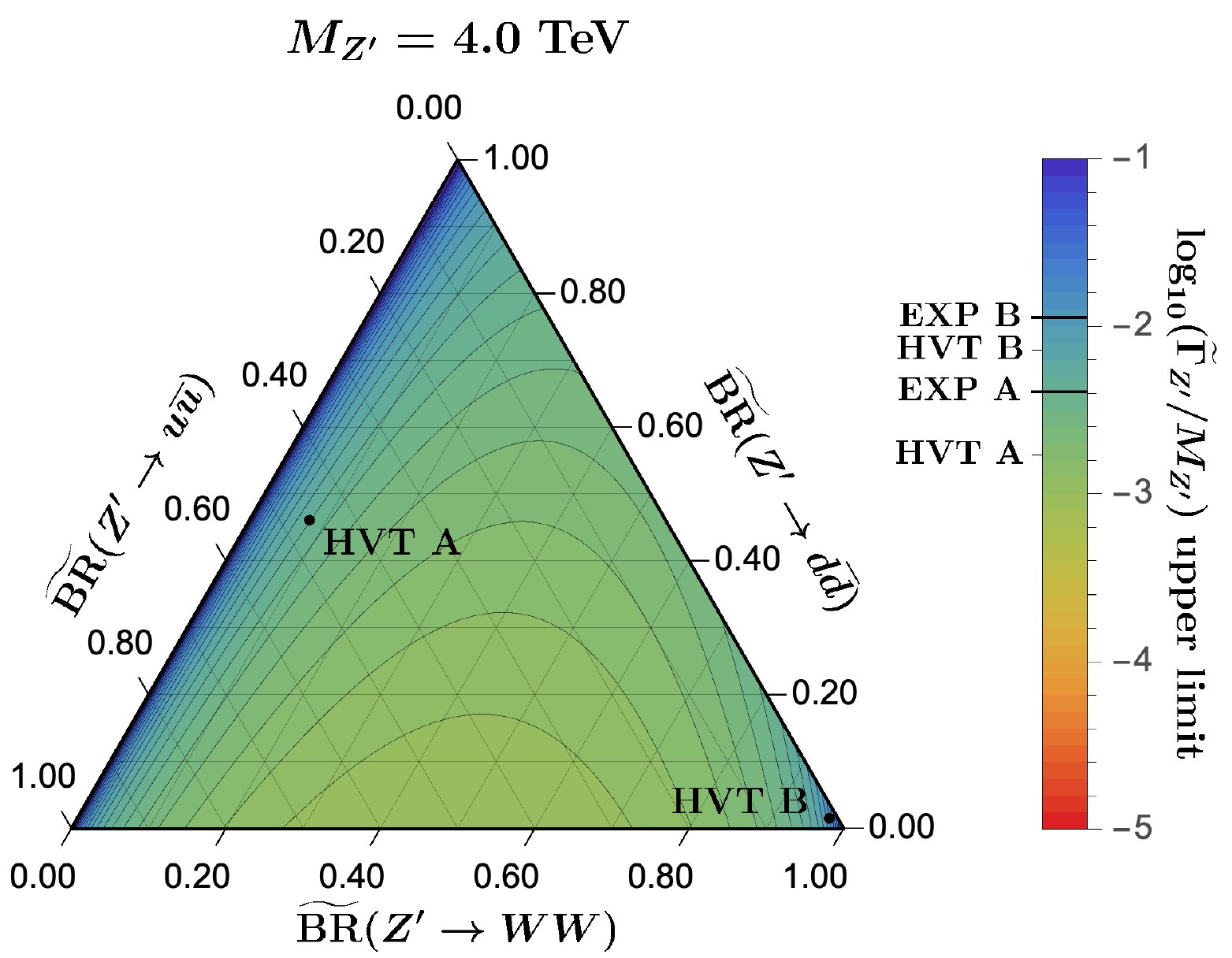}
    \caption{Ternary diagrams showing constraints on a neutral vector narrow resonance from DY production in the $WW$~\cite{Aad:2020ddw} and $ZH$ channels~\cite{Aad:2020tps}. The colors depict constraints on $\widetilde{\Gamma}_{Z^\prime} / m_{Z^\prime}$ for a selection of resonance masses. Points labeled ``HVT A'' and ``HVT B'' mark the locations in the parameter space corresponding to the BMs for model A and model B, respectively. The black lines through the legend (labeled ``EXP A'' and ``EXP B'') show the values of the experimental constraints on the BM HVT models, while the tick marks to the left of the legend (labeled ``HVT A'' and ``HVT B'') show their predicted values.}
    \label{fig:uuddWWspin1ternary}
  \end{figure}

  Conversely, while searches for a neutral vector resonance via DY production may not be ideal candidates for model-independent combined searches, one may instead probe its coupling to light quarks via ternary diagrams. We take for example the search for a DY-produced $Z^\prime$ resonance in the $WW$ channel, which dominates the constraints when compared to the $ZH$ channel for a majority of the HVT parameter space. The main production modes are via either $u \overline{u}$ or $d \overline{d}$ annihilation, and fig.~\ref{fig:uuddWWspin1ternary} shows constraints on $\widetilde{\Gamma}_{Z^\prime} / M_{Z^\prime}$ over a range of masses. The HVT BMs are also displayed on the ternary diagrams, both by points labeling their predicted branching ratios and by tick marks on the left of the legend labeling the predicted values of $\log_{10} ( \widetilde{\Gamma}_{Z^\prime} / M_{Z^\prime} )$. The solid lines through the legends label the experimental constraints on the BM points, so that one can more easily see that both BMs are excluded for $M_{Z^\prime} \le 3.0$~TeV, while they are both unconstrained at $M_{Z^\prime} = 4.0$~TeV. In these examples, the tilt in the contours of constant $\log_{10} ( \widetilde{\Gamma}_{Z^\prime} / M_{Z^\prime} )$ follows from the roughly 2-to-1 luminosity ratio of $u$-to-$d$ quarks in the proton's PDF.\footnote{For a discussion of $Z^\prime$ properties characterized in part by their couplings to quarks, see e.g. Ref.~\cite{Carena:2004xs}.}

  \subsection{Spin-2 Resonance}
  \label{sec:spin2}
  
  For a spin-2 resonance, we consider the production of a narrow resonance via gluon fusion in the $WW$ and $ZZ$ channels. Heavy spin-2 resonances are a generic feature of models of quantum gravity in extra dimensions, where Kaluza-Klein (KK) towers of heavy gravitons ($G_\textrm{KK}$) are predicted. Typical BM models for considering the lightest graviton mode are the original RS model, referred to as RS1~\cite{Randall:1999ee}, where all SM fields are localized on the IR brane, and the bulk RS model~\cite{Agashe:2007zd}, where the SM fields are allowed to propagate in the bulk. In RS1, the localization of the graviton in the warped bulk near the IR brane induces couplings to all SM fields which are only TeV suppressed, so the graviton has significant BRs to light fermions. On the other hand, in the bulk RS model light fields are localized toward the UV brane which greatly suppresses their couplings to gravitons. Instead, the graviton is mainly produced via gluon fusion or VBF and has significant BRs to $t \overline{t}$, $W W$, and $Z Z$. In what follows we will therefore consider the bulk RS model.

  \begin{figure}
    \centering
    \includegraphics[width=0.75\textwidth]{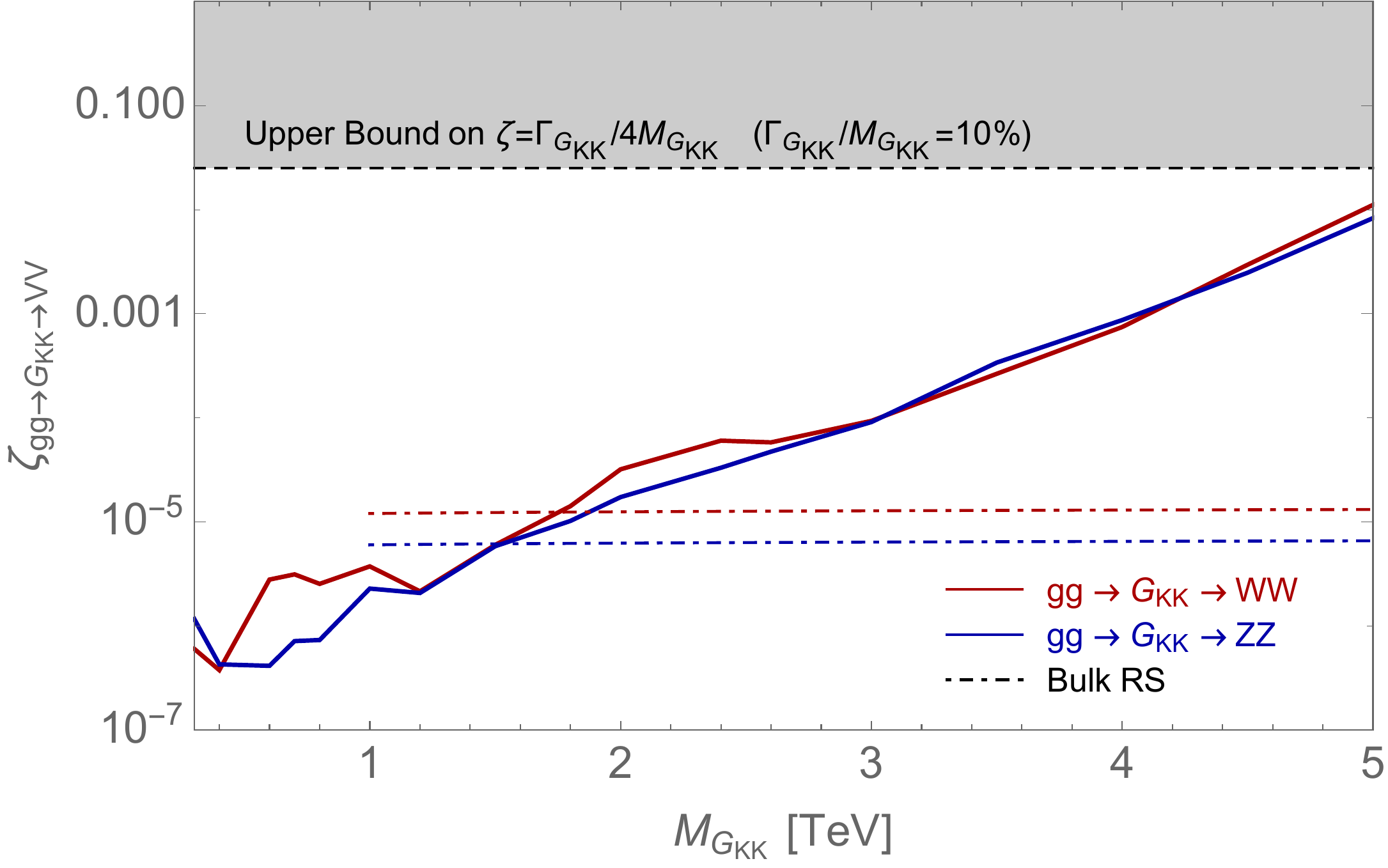}
    \caption{Constraints on spin-2 narrow resonance production from gluon fusion in the $WW$ and $ZZ$ channels~\cite{Aad:2020ddw}. Constraints from ATLAS are converted to upper limits on the $\zeta$ parameter, eq.~\eqref{eq:zeta}, represented by solid red and blue lines for the $WW$ and $ZZ$ channels, respectively. The gray shaded region corresponds to the upper limit on the product of branching ratios times $\Gamma_\phi / M_\phi = 10\%$, approximately where the NWA breaks down. The dot-dashed lines show the predictions from the Bulk RS BM model.}
    \label{fig:GGWWZZspin2zeta}
  \end{figure}

  The graviton's couplings to SM fields are determined by the parameter $k/\overline{M}_\textrm{Pl}$ where $k$ is the warped curvature scale and $\overline{M}_\textrm{Pl}$ is the reduced Planck mass. For our BM model of graviton production, we assume $k/\overline{M}_\textrm{Pl}=1.0$. Fig.~\ref{fig:GGWWZZspin2zeta} shows constraints from ATLAS searches in the $WW$ and $ZZ$ channels~\cite{Aad:2020ddw}, converted into the language of simplified limits. The constraints from each channel are competitive except for the region below $M_{G_\textrm{KK}} \lesssim 1.2$~TeV, where the $ZZ$ channel dominates. Also shown is the prediction from the bulk RS BM model, which sets a limit on the mass of the graviton of $M_{G_\textrm{KK}} \gtrsim 1.7~(1.5)$~TeV in the $WW$ ($ZZ$) channel.

  \begin{figure}
    \includegraphics[width=0.475\textwidth]{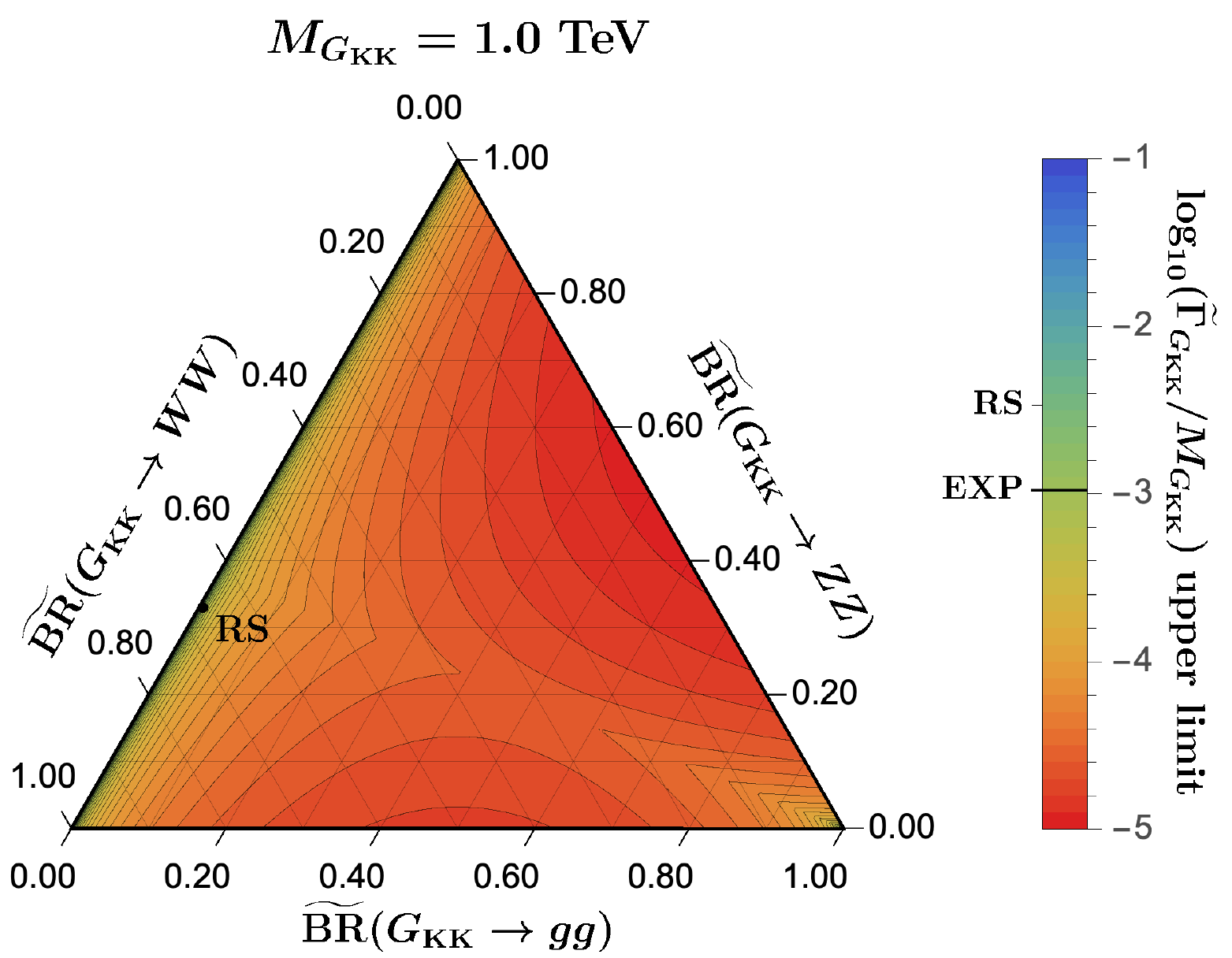}
    \hspace*{0.025\textwidth}
    \includegraphics[width=0.475\textwidth]{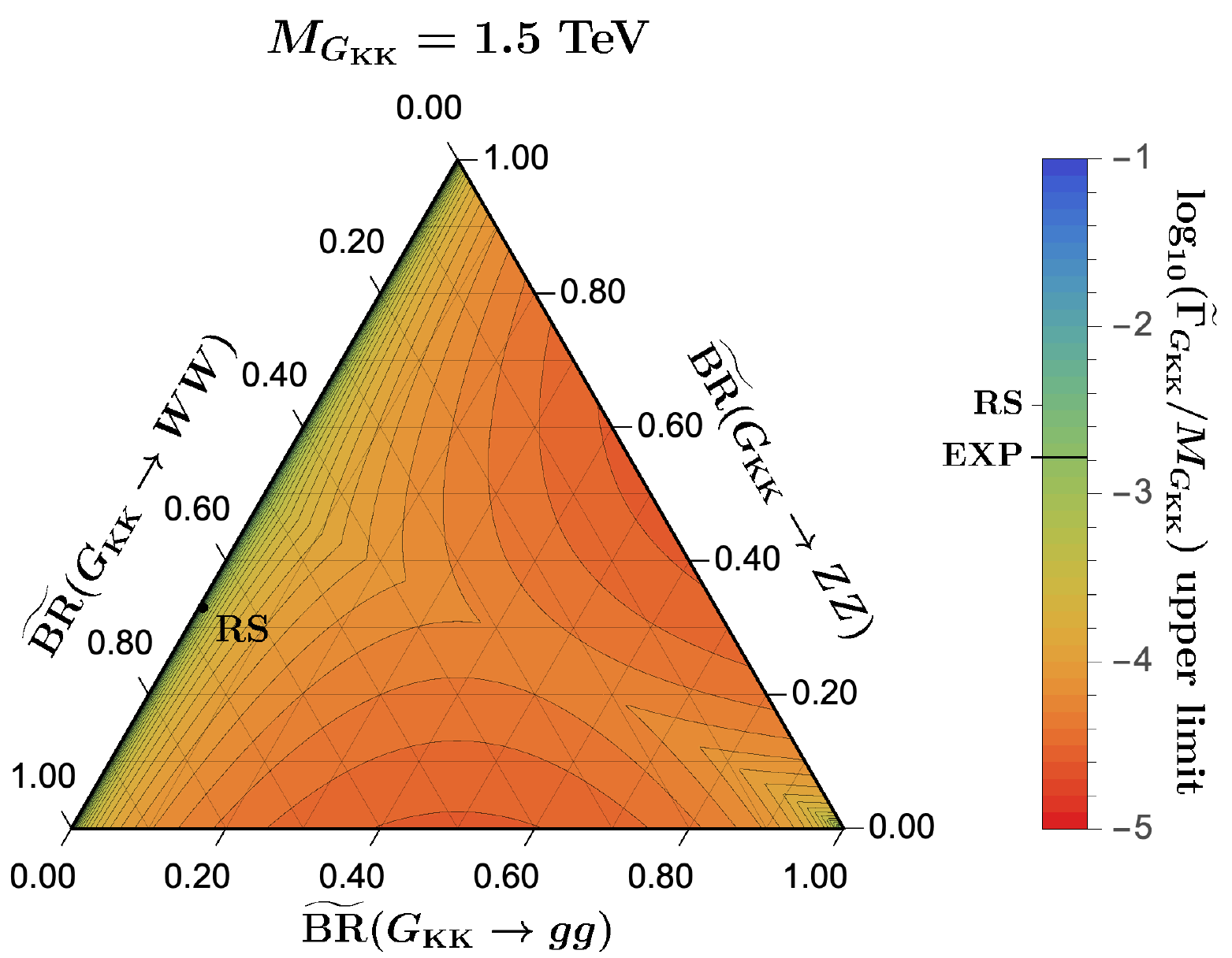}
    \vspace*{5mm} \\
    \includegraphics[width=0.475\textwidth]{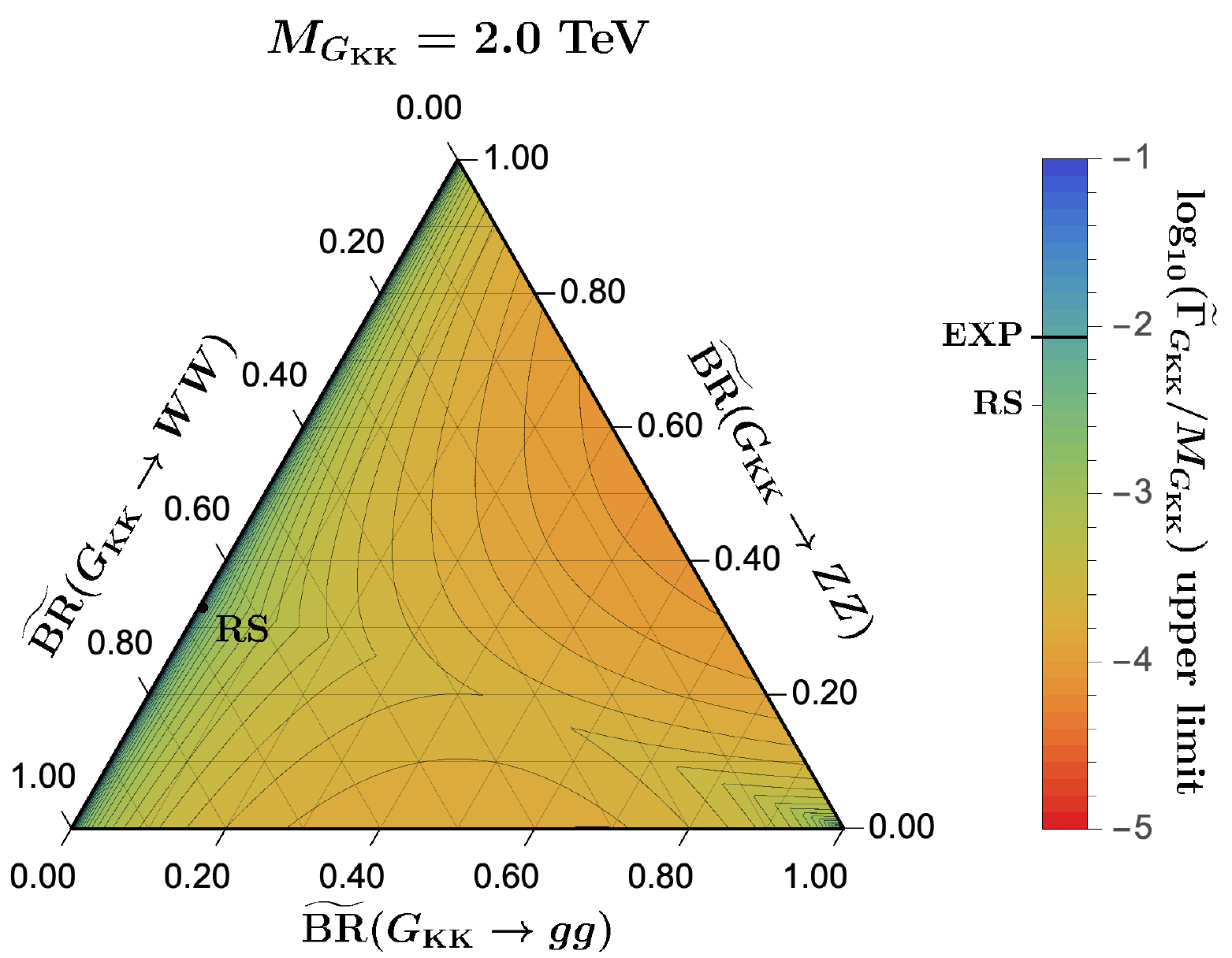}
    \hspace*{0.025\textwidth}
    \includegraphics[width=0.475\textwidth]{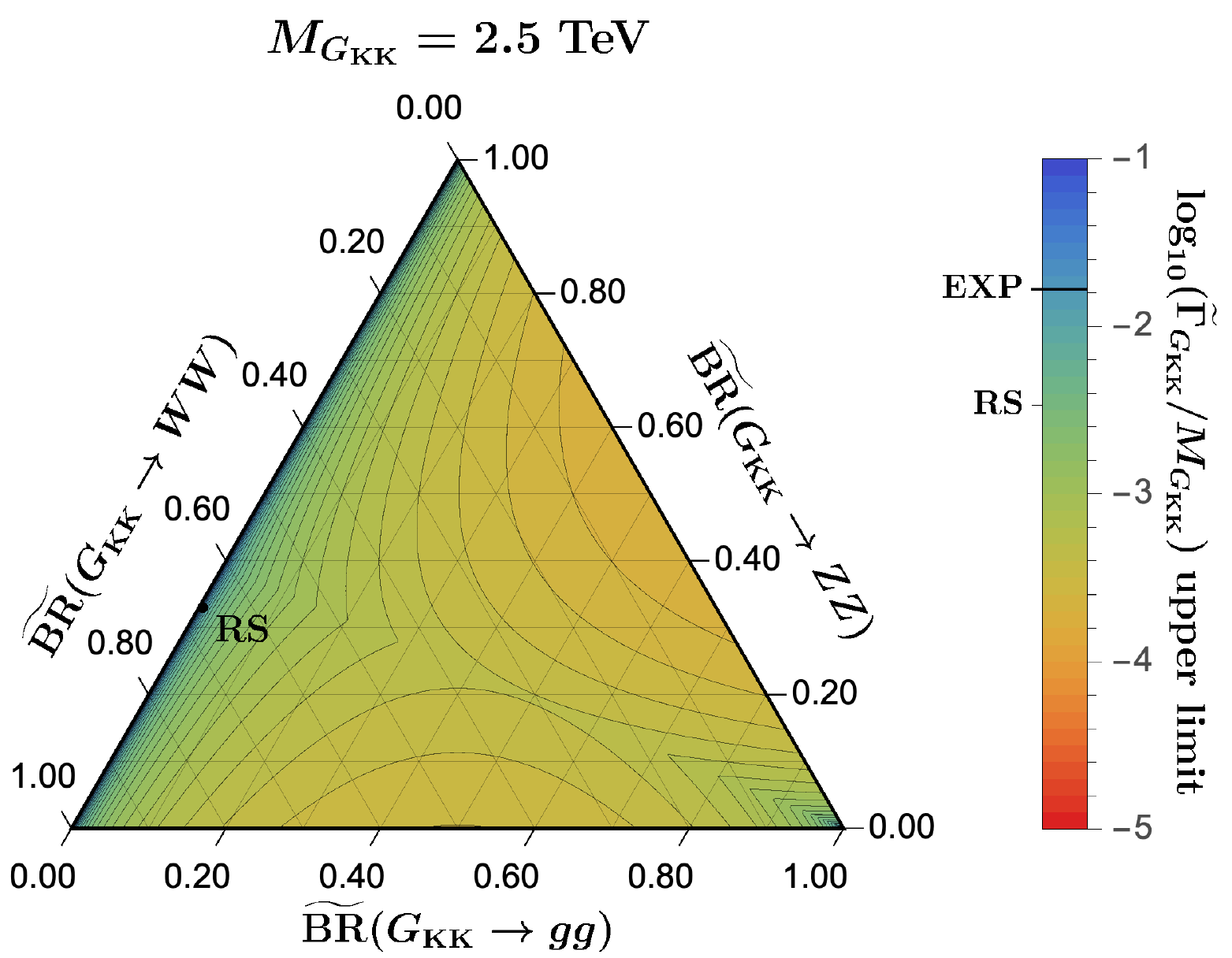}
    \vspace*{5mm}
    \caption{Ternary diagrams showing constraints on spin-2 narrow resonance production from gluon fusion in the $WW$ and $ZZ$ channels~\cite{Aad:2020ddw}. The colors depict constraints on $\widetilde{\Gamma}_R / m_R$ for a selection of resonance masses. The black dot represents the location in the parameter space corresponding to the bulk RS BM model. The black line through the legend labeled ``EXP'' shows the value of the experimental constraints on the bulk RS BM model, while the tick mark to the left of the legend labeled ``RS'' shows its predicted value.}
    \label{fig:GGWWZZspin2ternary}
  \end{figure}

  Fig.~\ref{fig:GGWWZZspin2ternary} shows ternary diagrams associated with these searches, displaying constraints on $\widetilde{\Gamma}_{G_\textrm{KK}} / M_{G_\textrm{KK}}$ for a selection of graviton masses. The predictions from the bulk RS BM are also shown, both by a point labeling their predicted branching ratios and by a tick mark on the left of the legend labeling the predicted value of $\log_{10} ( \widetilde{\Gamma}_{G_\textrm{KK}} / M_{G_\textrm{KK}} )$. The solid line through the legend labels the experimental constraint on the BM point, so that one can more easily see that the BM model is excluded for $M_{G_\textrm{KK}} \le 1.5$~TeV and unconstrained for $M_{G_\textrm{KK}} \ge 2.0$~TeV, in agreement with fig.~\ref{fig:GGWWZZspin2zeta}.

  \section{Further Generalization: N-Simplexes}
  \label{sec:simplex}
  
  In this section, we indicate how to further extend the simplified limits framework to situations in which more than a few branching ratios are important sources of experimental constraints on new resonances.  
 
  In sec.~\ref{sec:ternary}, we introduced a means of extending searches for narrow resonances into the multidimensional parameter space of BRs, focusing on situations where only three BR's were of greatest importance: either two production modes and one primary decay mode, or one production mode and a pair of experimentally distinguishable final states.  Section~\ref{sec:apps} explored a number of specific examples, showing how ternary diagrams can provide insight.
  
  Let us now consider the situation where a new resonance has N+1 branching ratios, all able to provide significant experimental constraints.  Then we can generalize the work of sec.~\ref{sec:ternary}, by noting that the branching ratios now obey the sum rule:
  \begin{align}
    \sum_{i = 1}^{N+1} \textrm{BR}_i = 1 \, ,
  \end{align}
  which tells us that the dimension of the space of independent BR's is N.  Just as compositional data in three variables can be represented by a ternary plot, where the ratios summing to 1 are plotted within a two-dimensional equilateral triangle, so compositional data in N+1 variables can be represented in a simplicial sample space where the ratios summing to 1 are located within an N-simplex.\footnote{A simplex generalizes a triangle to arbitrary dimensions; it is the minimal polytope in the space of a given number of dimensions. A 0-simplex is a point; a 1-simplex is a line segment; a 2-simplex is a triangle; a 3-simplex is a tetrahedron; a 4-simplex is a 5-cell, and so forth.}
  
  Moreover, if the new resonance has M+1 branching ratios, of which only N+1 are able to provide significant experimental constraints, we can generalize the notion of "effective" branching ratios from sec.~\ref{sec:ternary} as well.  Now, we have $\sum_{i = 1}^{N+1} \textrm{BR}_i = 1 - \sum_{i = N+2}^{M+1} \textrm{BR}_i$. We can thus define ``effective'' BRs,
  \begin{align}
    \widetilde{\textrm{BR}}_i \equiv \textrm{BR}_i \left ( 1 - \sum_{j = N+2}^{M+1} \textrm{BR}_j \right )^{-1} \, ,
  \end{align}
  which satisfy the unitary sum rule, $\sum_{i = 1}^{N+1} \widetilde{\textrm{BR}}_i = 1$, implicit in the construction of simplex diagrams. To interpret constraints displayed on the N-simplex diagrams in the framework of simplified limits, one must then also define an effective width,
  \begin{align}
    \widetilde{\Gamma}_R \equiv \Gamma_R \left ( 1 - \sum_{j = N+2}^{M+1} \textrm{BR}_j \right )^2 \, .
  \end{align}
  An N-simplex diagram with sides spanning the range $[0,\,1]$ in this context generically displays the space of effective BRs, where $\widetilde{\textrm{BR}} \ge \textrm{BR}$ and $\widetilde{\Gamma}_R \le \Gamma_R$ with the equalities saturated only when $N=M$.
  
  Again, within the simplified limits framework, this construction allows us to handle a wider variety of scenarios.  On the one hand, we can unfold the ambiguity introduced by deconvolving the hadronic PDFs when there is more than one production mode. On the other hand, we can combine search results for multiple experimentally distinguishable final states without introducing further model-dependent assumptions about the relationship between BRs. 
  
  While one cannot easily plot a higher-dimensional N-simplex as a two-dimensional image, one can nonetheless still perform a statistical analysis to discern how the experimental constraints shape the allowed region of the N-simplex and check whether a given new resonance's location within the N-simplex is in the allowed region.  One might even illustrate this in a journal article by displaying the ternary diagram sub-space of the full N-simplex that provides the strongest limit on the model in question.

  \section{Discussion}
  \label{sec:disc}
  
  In this article we have introduced a more integrative method of presenting constraints on production and decay of narrow resonances when multiple branching ratios yield valuable experimental information about the properties of new resonances.  The method utilizes the NWA to parameterize constraints in terms of products of BRs, the mass of the resonance, and the total width of the resonance.  We have seen that representing the results of searches for narrow resonances in terms of the parameterization of the simplified limits framework---$\widetilde{\textrm{BR}}$s, $\widetilde{\Gamma}_R$, and $m_R$---provides a natural context for combining the statistics from multiple search channels for a common resonance.   
  
  We have largely focused on cases where only three channels (two production and one decay mode, or vice versa) are relevant, by employing ternary diagrams to display the combined constraints on the expanded space of BRs.  We have illustrated applications to resonances of spin 0, spin 1, and spin 2, arising in a variety of beyond-the-standard-model scenarios.  Our approach clearly offers a more model-independent method of interpreting constraints from multiple channels compared to the traditional product of production cross section times BR, which would require making assumptions about the relationship between decay BRs. It is also applicable to situations with multiple production modes by unfolding the uncertainty inherent in the one-dimensional simplified limits parameter $\zeta$. This method is complementary to traditional limits, with $\sigma \times \textrm{BR}$ offering the cleanest display of constraints at the expense of sometimes introducing specific model assumptions, while ternary diagrams can encompass more a more model-independent parameter space, easily translatable to a variety of disparate models.  
  
  While the applications we considered in detail had three-channel parameter spaces, we have also discussed how the use of ternary diagrams can be readily generalized to the use of N-simplex diagrams for situations where additional branching ratios also offer valuable experimental constraints on the properties of new resonances.  Making use of this more generalized method will, of course, rely on the availability of data about the multiple branching ratios.  Indeed, having access to digital data sets for searches combining multiple experimental channels would be ideal for documenting and leveraging limits on a many-dimensional parameter space, capable of encompassing all of the initial and final state BRs relevant for the searches being considered. Recently, there has been a tremendous effort to make digitalized data available to all researchers with the introduction of the HepData repository~\cite{Maguire:2017ypu}. 
  
  As a closing thought, we would like to advocate for experimental collaborations to provide these larger digital data sets (parameterized in terms of the BRs, total width, and mass of the new resonance) to supplement the information presentable in the traditional article format. This will enable the data to be most fully leveraged to explore the widest possible range of models in detail, enabling constraints to be quickly understood for a plethora of interesting theories.

  \section{Acknowledgments}
  \label{sec:ack}
  
  We thank J. Duarte for suggesting the application of this work to digital data repositories. We also thank D. Foren, K.A. Mohan, D. Sengupta, and X. Wang for useful discussions and comments. This material is based upon work supported by the National Science Foundation under Grant No.~PHY-1915147. P.I. was supported by the CUniverse research promotion project of Chulalongkorn University in Bangkok, Thailand, under Grant No.~CUAASC.

  
  \bibliography{mybibfile}
  \bibliographystyle{JHEP}
  
      
\end{document}